# Rebooting Internet Immunity

*Gregory M. Dickinson\**

## Abstract

*We do everything online. We shop, plan travel, invest, socialize, and even hold garage sales. Even though we may not care whether a company operates online or in the physical world—the distinction has important consequences for the companies themselves. Online and offline entities are governed by different rules. Under Section 230 of the Communications Decency Act, online entities—but not physical-world entities—are immune from lawsuits related to content authored by their users or customers. As a result, online entities have been able to avoid claims for harms caused by their negligence and defective product designs simply because they operate online.*

*The reason for the disparate treatment is the internet's dramatic evolution over the last two decades. The internet of 1996 served as an information repository and communications channel and was well governed by Section 230, which treats internet entities as another form of mass media. Because Facebook, Twitter, and other online companies could not possibly review the mass of content that flows through their systems, Section 230 immunizes them from claims related to user content. But content distribution is not the internet's only function, and it is even less so now than it was in 1996. The internet also operates as a platform for the delivery of real-world goods and services and requires a correspondingly diverse immunity doctrine. This Article proposes refining online immunity by limiting it to claims that threaten to impose a content-moderation burden on internet defendants. Where a claim is preventable other than by content moderation—for example, by redesigning an app or website—a plaintiff could freely seek relief, just as in the physical world. This approach empowers courts to identify culpable actors in the virtual world and treat like conduct alike wherever it occurs.*

## Table of Contents

---

\* Assistant Professor, St. Thomas University College of Law; Fellow, Stanford Law School, Law, Science, and Technology Program; J.D., Harvard Law School; B.S., Computer Science, Houghton College. The author represented members of the U.S. Congress in a case before the Wisconsin Supreme Court on matters related to the topic of this Article. The views expressed are solely those of the author. For their insights and generous comments, thanks to John Goldberg, Thomas Kadri, Dmitry Karshtedt, Kate Klonick, Kyle Langvardt, Mark Lemley, Irina Manta, Agnieszka McPeak, Alan Rozenshtein, Chas Tyler, Eugene Volokh, and participants at the Junior Scholars Workshop on Tech-Company Governance at the University of Minnesota Law School in December 2019 and the Legal Research in Progress Workshop at Stanford Law School in May 2020.

## INTRODUCTION

The internet has transformed nearly every facet of human activity, from how we learn, work, and communicate to how we shop, plan travel, invest, socialize, and even hold garage sales. The virtual world has become so expansive that it now operates in nearly perfect parallel to the real world, with a specialized website or app available to accomplish any physical-world task imaginable. Looking to round out your summer wardrobe or track down the latest smart watch? Head to the local shopping mall. Or to Amazon.com—either will work. Need to buy or sell secondhand kids' clothes? Try the local flea market, or eBay, or, if you are feeling fancy, Poshmark. The internet is now so diversely populated that we scarcely notice and hardly care whether the entities with which we interact operate in the physical or virtual world.

But the question has dramatic consequences for the entities themselves. Even when they engage in exactly the same conduct, online and offline entities are subject to completely different legal regimes. Unlike their physical-world counterparts, online entities are immune from many types of state and federal lawsuits—specifically, those that relate to content created by third parties—under Section 230[1] of the Communications Decency Act ("CDA") of 1996.[2] As a result, online entities have been able to operate unlawful businesses—facilitation of prostitution and unlawful gun sales to name just two—free from the threat of civil liability that they would face in the physical world.[3] And even those online entities with more salutary business models enjoy a competitive advantage over their physical-world counterparts. Immunity from lawsuits means lower litigation costs and fewer judgments and damage awards for plaintiffs.

Why bar behavior in the physical world only to allow that same behavior in the virtual world, where it may be even easier to accomplish? The question sounds simple enough. It is one of law's central tenets, after all, that like cases be treated alike. Even so, few topics rouse such passion or touch so many ongoing debates as the question of internet immunity. Every day brings fresh controversy, call for change, or proposed legislation. From cyberbullying,[4] online governance,[5] and freedom of expression,[6] to big-tech antitrust concerns,[7] pri-vacy,[8] and tort liability,[9] the effects of internet immunity law are as wide ranging as the internet itself.

---

[1] 47 U.S.C. § 230.

[2] Pub. L. No. 104-104, tit. V, 110 Stat. 56, 133–43 (codified in scattered sections of 18 and 47 U.S.C.).

[3] *See infra* Sections I.A–.B; *see also, e.g.*, Alina Selyukh, *Section 230: A Key Legal Shield for Facebook, Google Is About To Change*, NPR (Mar. 21, 2018, 5:17 PM), https://www.npr.org/sections/alltechconsidered/2018/03/21/591622450/section-230-a-key-legal-shield-for-facebook-google-is-about-to-change [https://perma.cc/27L8-44GM] (discussing how Section 230 shielded Backpage from civil liability for facilitation of child sex trafficking).

[4] *See* Erica Goldberg, *Free Speech Consequentialism*, 116 COLUM. L. REV. 687, 744–45 (2016) (noting that current internet immunity doctrine bars claims against online entities for revenge porn and other forms of cyberbullying); Andrew Gilden, *Cyberbullying and the Innocence Narrative*, 48 HARV. C.R.-C.L. L. REV. 357, 389–90 (2013) (critiquing proposals to narrow online immunity to protect gay teens from harassment on ground that such efforts obscure the power of individual agency).

[5] *See* Jack M. Balkin, *Free Speech in the Algorithmic Society: Big Data, Private Governance, and New School Speech Regulation*, 51 U.C. DAVIS L. REV. 1149, 1182–93 (2018) (discussing platforms' role as regulators of free speech in digital era); Jennifer Daskal, *Speech Across Borders*, 105 VA. L. REV. 1605, 1637–44 (2019) (discussing geographic scope of online platforms' content-filtering determinations and implications for territorial sovereignty); Kate Klonick, *The New Governors: The People, Rules, and Processes Governing Online Speech*, 131 HARV. L. REV. 1598, 1599–1613 (2018) (tracing the ability of private platforms like Facebook to make content-moderation decisions regarding user-submitted content to Section 230); Frank Pasquale, *Two Narratives of Platform Capitalism*, 35 YALE L. & POL'Y REV. 309, 316–19 (2016) (offering two possible narratives of the distributed online platform and implications for each on regulatory and self-governance policy decisions); *see also* David R. Johnson & David Post, *Law and Borders—The Rise of Law in Cyberspace*, 48 STAN. L. REV. 1367, 1367 (1996) (arguing just prior to Section 230's enactment that internet regulation would require its own distinct principles); Lawrence Lessig, Commentary, *The Law of the Horse: What Cyberlaw Might Teach*, 113 HARV. L. REV.

---

501, 502 (1999) (arguing that the study of cyberlaw can illuminate principles that affect the real world).

6  *See* Danielle Keats Citron, *Extremist Speech, Compelled Conformity, and Censorship Creep*, 93 NOTRE DAME L. REV. 1035, 1036–40 (2018) (exploring departure of online platforms from U.S. First Amendment values and dangers of bowing to international pressure to self-regulate); Danielle Keats Citron & Helen Norton, *Intermediaries and Hate Speech: Fostering Digital Citizenship for Our Information Age*, 91 B.U. L. REV. 1435, 1453–84 (2011) (noting that Section 230 insulates platforms from legal liability and offering proposals for online platforms to voluntarily respond to online hate speech); Eric Goldman, *Why Section 230 Is Better Than the First Amendment*, 95 NOTRE DAME L. REV. REFLECTION 33, 36–46 (2019), https://scholarship.law.nd.edu/ndlr_online/vol95/iss1/3 [https://perma.cc/53MP-4GQU] (discussing Section 230's enhanced substantive and procedural protections for online entities beyond those of the First Amendment); Rebecca Tushnet, *Power Without Responsibility: Intermediaries and the First Amendment*, 76 GEO. WASH. L. REV. 986, 1009 (2008) (arguing that Section 230 immunity should include a corresponding limit on an intermediary's ability to censor speech); Felix T. Wu, *Collateral Censorship and the Limits of Intermediary Immunity*, 87 NOTRE DAME L. REV. 293, 295–96 (2011) (noting speech-enhancing effects of Section 230 due to its preventing imposition of liability on intermediaries for harmful or offensive speech that they might otherwise censor).

7  *See* C. Scott Hemphill, *Disruptive Incumbents: Platform Competition in an Age of Machine Learning*, 119 COLUM. L. REV. 1973, 1974–93 (2019) (identifying potential sources of competition among dominant participants in online platform market and offering proposals to maximize competition); Lina M. Khan, *The Separation of Platforms and Commerce*, 119 COLUM. L. REV. 973, 1037–92 (2019) (proposing bars on entities' engaging in new lines of business as a check on dominance of small number of tech firms); Lina M. Khan & David E. Pozen, *A Skeptical View of Information Fiduciaries*, 133 HARV. L. REV. 497, 527–28 (2019) (noting Google and Facebook's capture of digital advertising market in United States and resultant effects on traditional publishing industry).

8  *See* Danielle Keats Citron, *Sexual Privacy*, 128 YALE L.J. 1870, 1952–53 (2019) (proposing modification to Section 230 immunity to spur platforms to action to protect against revenge porn and other invasions of sexual privacy); Bobby Chesney & Danielle Citron, *Deep Fakes: A Looming Challenge for Privacy, Democracy, and National Security*, 107 CALIF. L. REV. 1753, 1755–59, 1795–804 (2019) (describing rising danger to privacy and security posed by advances in technology for creating deep fakes and noting that Section 230 limits legal recourse against online entities that distribute such fakes).

9  *See* Ann Bartow, *Internet Defamation as Profit Center: The Monetization of Online Harassment*, 32 HARV. J.L. & GENDER 383, 384 (2009) (tracing rise of commercial reputation defense services to lack of traditional avenues of recourse to respond to online harassment); Danielle Keats Citron, *Mainstreaming Privacy Torts*, 98 CALIF. L. REV. 1805, 1836–44 (2010) (sketching vision for new era of privacy law and noting barrier that Section 230 poses to tortious enablement claims against online entities); Danielle Keats Citron & Benjamin Wittes, *The Problem Isn't Just Backpage: Revising Section 230 Immunity*, 2 GEO. L. TECH. REV. 453, 455–56 (2018) (proposing that online immunity be narrowed to allow claims against online entities that do not take reasonable steps to address unlawful third-party content); Benjamin Edelman & Abbey Stemler, *From the Digital to the Physical: Federal Limitations on Regulating Online Marketplaces*, 56 HARV. J. ON LEGIS. 141, 143 (2019) (noting bar that Section 230 poses to regulation of modern online marketplaces); Olivier Sylvain, *Intermediary Design Duties*, 50 CONN. L. REV. 203, 203 (2018) (suggesting that online immunity doctrine be updated to consider manner in which online entities elicit and use their users' data).

The reason that internet immunity reform rouses such passion is because the stakes are so very high. Since its enactment more than two decades ago, Section 230 has been a resounding success. Its broad protections against lawsuits related to third-party content shield online entities from an economically crippling duty to review the nearly inconceivable volume of data that flows through their systems.[10] Without such protection, online platforms might be compelled to censor user speech or disallow online posting altogether to avoid the risk of liability. Section 230's protections have been crucial to decades of free speech advances built on inexpensive and free flowing internet publishing technologies.[11]

Despite its nearly sacred status in the tech industry, however, Section 230 started off humbly,[12] and suffers from a humble problem: The Congress of 1996 did not foresee the internet of 2020, and the statute is now outdated. Section 230 assumes a publication-industry-like model of the internet—it encourages censorship and speaks in terms of "publisher[s] or speaker[s]" and "content provider[s]"[13]—and is well suited to govern the internet's information repositories and communications channels.[14] Online actors in this publication-centric internet model can be handily divided into three camps: content authors,[15] computer services[16] that provide access to content, and the internet users who consume the content.[17] With the virtual world so neatly divisible, potential wrongdoers are easy to identify. Any wrongdoing can be attributed to its active participants—the internet content creators who author content—not the passive computer services and their users who merely provide access to and view that content. With Section 230, Congress federalized the principle by adopting a bright line rule that none but an "information content provider" can be held liable for internet content.[18]

---

[10] *See infra* note 91 and accompanying text.

[11] *See infra* notes 92–94 and accompanying text.

[12] Section 230 was enacted as part of the much more comprehensive CDA and received little fanfare. Few media outlets included any discussion of Section 230 in their coverage of the CDA. *See* Jeff Kosseff, The Twenty-Six Words That Created the Internet 66–68 (2019).

[13] 47 U.S.C. § 230(c).

[14] *See infra* Section I.C.3.

[15] Content authors or "information content providers" are defined by Section 230 as "any person or entity that is responsible, in whole or in part, for the creation or development of information provided through the Internet or any other interactive computer service." 47 U.S.C. § 230(f)(3).

[16] Section 230 defines an "interactive computer service" as "any information service, system, or access software provider that provides or enables computer access by multiple users to a computer server, including specifically a service or system that provides access to the Internet and such systems operated or services offered by libraries or educational institutions." *Id.* § 230(f)(2). The archetypical computer services at the time of Section 230's enactment were the internet service providers Prodigy, CompuServe, and, later, America Online. *See* Kosseff, *supra* note 12, at 36.

[17] 47 U.S.C. § 230(c)(1).

But publication is not the internet's exclusive function, and that is even more the case now than it was in 1996. Section 230 now presides over a much different internet than the one it was designed to govern. The internet of today is much more diverse, specialized, and interactive. Of course, it continues to include many online platforms that transmit and host third-party content, but it also supports the delivery of innumerable real-world goods and services that would have been unimaginable twenty years ago.[19] Authoring or failing to moderate content flowing through their services is not the only way that online entities can cause harm.[20] Consequently, Section 230's bright line rule relying on content authorship as the deciding factor for immunity is poorly tailored for the internet that exists today.

Recognizing the growing problem, legal scholars and lawmakers of both political parties have proposed numerous reforms,[21] such as amending Section 230 to withhold immunity if online entities intentionally or knowingly facilitate illegal conduct,[22] profit from unlawful content,[23] improperly monetize user data,[24] or fail to abide by best practices in policing the online behaviors of their users.[25] Internet im-

---

[18] *See* 47 U.S.C. § 230(c); Zeran v. Am. Online, Inc., 129 F.3d 327, 330 (4th Cir. 1997) ("By its plain language, § 230 creates a federal immunity to any cause of action that would make service providers liable for information originating with a third-party user of the service.").

[19] *See infra* Section I.C. *See generally* Edelman & Stemler, *supra* note 9, at 143–51 (discussing online marketplaces that facilitate a variety of commercial transactions).

[20] *See infra* Section II.B.

[21] For a more detailed discussion of the various reform proposals, see *infra* Section III.A.1.

[22] *See* U.S. Dep't of Just., Section 230—Nurturing Innovation or Fostering Unaccountability? 14–15 (2020), https://www.justice.gov/file/1286331/download [https://perma.cc/6AU2-SKNN] [hereinafter DOJ Section 230 Recommendations] (recommending a "Bad Samaritan" carve out to remove immunity where online entities purposefully solicit third parties to engage in unlawful activities through their platforms); Citron & Wittes, *supra* note 9, at 455–56 (proposing denial of immunity if an entity fails to take reasonable steps to protect against known illegal activity).

[23] *See* Agnieszka McPeak, *Platform Immunity Redefined*, 62 Wm. & Mary L. Rev. (forthcoming 2021) (recommending that online immunity doctrine should incorporate principles of joint enterprise liability).

[24] *See* Sylvain, *supra* note 9, at 208–14 (discussing entities' intentional solicitation and sale or use of user data).

[25] *See* Danielle Keats Citron & Benjamin Wittes, *The Internet Will Not Break: Denying Bad Samaritans § 230 Immunity*, 86 Fordham L. Rev. 401, 419 (2017) (recommending that immunity be contingent on an entity's reasonable efforts to prevent unlawful uses of its service). Another set of recent proposals would modify Section 230 not because it is out of date, but to

munity reform continues to face significant opposition, however, from the tech industry and from those legal scholars[26] who worry that changes to Section 230 could do more harm than good. Changes to the statute could curtail freedom of expression on the internet and spur online platforms to censor user speech by undermining the critical protection Section 230 provides against lawsuits for failing to moderate third-party content.

To navigate those competing concerns, this Article proposes a carefully tailored refinement to internet immunity doctrine that would expressly bar any claim that would impose a content-moderation burden on an internet platform but would allow other claims to proceed. Thus, where an alleged harm is preventable by a means other than content moderation—for example, by redesigning an app or website—a plaintiff could freely seek relief, just as in the physical world. By shifting the internet-immunity inquiry from the publication-focused question of content authorship to the more generally applicable question of content moderation, this approach would arm courts with a more flexible analytical tool and free them from Section 230's outdated publication-focused model. The approach would empower courts to eliminate the online-versus-offline disparity and treat like conduct alike in both the physical and virtual world

This Article approaches the problem in three parts. Part I uses a mass shooting wrongful death case decided by the Wisconsin Supreme Court as a case study to analyze the disparity in the law governing internet versus real-world defendants and to illustrate the harms that can flow from immunizing certain online behavior. Part II examines various contexts in which Section 230 is poorly suited to govern the modern, heterogenous internet. Finally, Part III presents a new frame-

---

use the threat of its removal to pressure entities into action to address concerns about large tech companies' concentrated power over speech and commerce. *See* Online Freedom and Viewpoint Diversity Act, S. 4534, 116th Cong. § 2 (2020) (removing liability protection where an online entity moderates third-party content unless it does so with an "objectively reasonable belief" that the content is obscene, lewd, lascivious, filthy, excessively violent, harassing, "promoting self-harm, promoting terrorism, or unlawful"); Ending Support for Internet Censorship Act, S. 1914, 116th Cong. (2020) (prohibiting content moderation from politically biased standpoint, to be enforced by Federal Trade Commission ("FTC") audit confirming neutral censorship practices as condition of immunity); *see also* Preventing Online Censorship, Exec. Order No. 13925, 85 Fed. Reg. 34,079, 34,080 (May 28, 2020) (stating that Section 230 does not immunize online "behemoths . . . when they use their power to censor content and silence viewpoints that they dislike" and directing the Federal Communications Commission ("FCC") to consider rulemaking that would interpret Section 230 to immunize entities from claims related to content-moderation decisions only when they fall within Section 230(c)(2)'s Good Samaritan provision).

26 Goldman, *supra* note 6, at 36–46 (describing Section 230's substantive and procedural advantages to online entities and opposing reform efforts that could affect those advantages).

work for evaluating an online entity's entitlement to immunity, explores some of the obstacles to reform, and discusses how this Article's proposed framework attempts to navigate those obstacles.

## I.   THE ONLINE-VERSUS-OFFLINE DISPARITY

This Part uses a 2012 mass shooting as a case study to illustrate the disparity in law governing online and offline entities and to set the stage for discussion in Parts II and III of the sources, consequences, and potential solutions for the disparity.

### A.   *Case Study: The 2012 Azana Spa Shootings*

Zina Daniel Haughton was abused by her husband for more than a decade.[27] On Wednesday, October 3, 2012, Zina decided to change that.[28] Following yet another assault, Zina called 911, and under the protection of police, she moved out of the home she shared with her husband, Radcliffe Haughton.[29] The next Monday, Zina filed a petition[30] seeking a restraining order against her husband, which the court granted, following a fifty-minute hearing.[31] The hearing included testimony from Zina that Radcliffe had beaten her, choked her until she could not breathe, and "terrorize[d] [her] every waking moment."[32] The judge issued an order prohibiting Radcliffe from having any contact with Zina.[33] Based on a finding of "clear and convincing evidence" that Zina's husband might use a firearm to harm Zina, the court order also prohibited Zina's husband from possessing a firearm at any time over the next four years, the maximum time permitted under Wisconsin law.[34]

Although Zina did everything the system asked of her, any feeling of safety was short lived. Knowing he would be barred from purchasing a firearm through any licensed dealer, Zina's husband shopped for firearms available from private sellers on Armslist.com.[35]

---

[27] Complaint at ¶¶ 1, 28, Daniel v. Armslist, LLC, No. 2015CV008710 (Milwaukee Cty. Wis. Cir. Ct. Nov. 20, 2015) [hereinafter Daniel Complaint].

[28] *Id.* at ¶¶ 2, 26.

[29] *Id.*

[30] *See* Petition for Temporary Restraining Order, Petitioner v. Haughton, No. 2012FA006234 (Milwaukee Cty. Wis. Cir. Ct. Oct. 8, 2012).

[31] *See* Injunction & Order, Petitioner v. Haughton, No. 2012FA006234 (Milwaukee Cty. Wis. Cir. Ct. Oct. 18, 2012); Daniel Complaint, *supra* note 27, at ¶¶ 30, 33.

[32] Daniel Complaint, *supra* note 27, at ¶¶ 3, 31.

[33] *Id.* at ¶¶ 3, 33.

[34] *Id.* at ¶¶ 3, 33–34.

[35] *Id.* at ¶¶ 83–91. Circumvention of firearms restrictions through an online-facilitated private transactions is a common problem. *See* Dan Frosch & Zusha Elinson, *Man Pleads Guilty in*

The day after the judge's order, Radcliffe browsed Armslist.com to identify two firearms for sale nearby, an AK-47 assault rifle and an FNP-40 semiautomatic pistol.[36] He contacted the sellers first through the Armslist.com website and then by phone to arrange a meeting.[37] Radcliffe purchased the FNP-40 pistol and three fourteen-round magazines in a cash transaction completed in the front seat of a car in a McDonald's parking lot.[38]

On the morning of Monday, October 21, 2012, just three days after Zina obtained a restraining order against him, Radcliffe walked into the salon where Zina and her daughter worked, armed with the pistol he had purchased a day before.[39] As he shouted for everyone to "Get down!" and those in the salon dove to the floor, Zina calmly approached and pleaded with her husband not to hurt anyone.[40] Radcliffe began shooting. He fired first at Zina; as she lay on the floor bleeding, Radcliffe began shooting her coworkers.[41] After stalking through the salon firing at its employees and customers, Radcliffe returned to his wife to shoot her in the head, killing her.[42] Zina's daughter, Yasmeen, was there to witness the horror.[43] Yasmeen survived only because of the bravery of a coworker who positioned herself in front of Yasmeen as Radcliffe took aim, losing her own life in the process.[44] All told that day, Radcliffe killed Zina and two other women at the spa, and injured four others before retreating to the second floor and taking his own life.[45]

Following the tragedy, Yasmeen commenced a lawsuit alleging numerous causes of action against Armslist.com, including negligence, public nuisance, wrongful death, and aiding and abetting tortious con-

---

*Gun Sale to Mass Shooter*, WALL ST. J., Oct. 8, 2020, at A6 (describing rising number of individuals who sell guns online, as a business, to individuals they meet through websites like Armslist.com, but who fail to conduct the background checks that are required of frequent private sellers).

[36] Daniel Complaint, *supra* note 27, at ¶¶ 4–5, 83–91, 101.

[37] *Id.* at ¶¶ 91–95, 101.

[38] *Id.* at ¶¶ 91–102. The purchase of the AK-47 rifle was never completed. The seller became skittish after Radcliffe insisted that he needed the gun immediately and could not make the purchase on Sunday because he would be in church at all times. *Id.* at ¶ 101.

[39] *Id.* at ¶¶ 111–12.

[40] *Id.* at ¶¶ 113–14; Carlos Sadovi, *Witness Calls Suspect's Slain Wife a Hero*, CHI. TRIB., Oct. 23, 2012, at 6.

[41] Daniel Complaint, *supra* note 27, at ¶¶ 114–23.

[42] *Id.* at ¶ 124.

[43] *Id.* at ¶ 118.

[44] *Id.* at ¶¶ 118–19.

[45] *Id.* at ¶¶ 126–27; Ryan Haggerty, Carlos Sadovi & Lisa Black, *Spa Shooter Had Threatened Wife*, CHI. TRIB., Oct. 23, 2012, at 6.

duct.[46] The claims were aggressive; they stretched the boundaries of Wisconsin tort law. Yet the Wisconsin trial court never considered whether Yasmeen had pleaded any cognizable tort claims. Instead, without written decision, the court granted Armslist.com's motion to dismiss pursuant to Section 230 of the CDA on the ground that federal law preempts all claims against websites that provide communication tools that have neutral and lawful purposes.[47] The fundamental questions the suit raised about Wisconsin tort law were thus deemed moot and never considered.[48]

Dismissal might have been the right outcome. Armslist.com, like a gun-show organizer, merely facilitates firearms sales between other individuals. Such entities are separated from subsequent wrongful conduct by at least two volitional actors: the private gun sellers and the buyers who pull the trigger. But consider how different the analysis is in the physical world.

## B. Divergent Physical and Virtual Worlds

### 1. The Physical World

To see the point, imagine a worst-case scenario in the physical world. "Illegal Gun Sales Warehouse" opens a gun shop designed "to facilitate and profit from the illegal purchase of firearms by people who are prohibited from purchasing" them.[49] The store "deliberately refuses to employ reasonable care to prevent foreseeable harm to others" and "[i]nstead . . . is designed to cater to illegal purchases."[50] At the Illegal Gun Sales Warehouse entrance, a clerk directs customers to a room where they can purchase guns in the ordinary fashion, including background check and ID requirements. But the clerk also informs customers that, in the back of the store, there is a private room where the Illegal Gun Sales Warehouse hosts third-party private sellers, who are well known to make anonymous sales, no questions asked, to anyone with enough cash.

---

[46] Daniel Complaint, *supra* note 27, at ¶¶ 128–208.

[47] *See* Daniel v. Armslist, LLC, 926 N.W.2d 710, 716 (Wis. 2019) (discussing result in trial court), *cert. denied*, 140 S. Ct. 562 (2019).

[48] Following dismissal of the action, the plaintiff appealed to Wisconsin's intermediate appellate court, the Court of Appeals of Wisconsin, which reversed the trial court's determination and reinstated the action. Daniel v. Armslist, LLC, 913 N.W.2d 211, 224 (Wis. Ct. App. 2018), *rev'd*, 926 N.W.2d 710. Armslist requested review by the Wisconsin Supreme Court, which it granted, reversed the appellate court's decision, and dismissed the action as barred by Section 230. *Daniel*, 926 N.W.2d at 727.

[49] *See* Brief of Plaintiff-Appellant Yasmeen Daniel at 1–2, *Daniel*, 926 N.W.2d 710 (No. 2017AP344). Thanks to Jon Lowy and the Brady Center to Prevent Gun Violence from whose appellate brief on behalf of Plaintiff Yasmeen Daniel this example is drawn.

[50] *Id.* at 1.

Such a case again presents complex questions of tort law. What duties, if any, does an entity that facilitates private gun sales between third parties have to potential victims of the gun buyers that they help to pair with gun sellers? When is an act of gun violence following an illegal sale reasonably foreseeable? If an act is foreseeable, when does the intentionally wrongful conduct of the shooter disrupt the chain of causality?

In the physical world, courts considering claims against private gun sale facilitators, such as gun show promoters, employ the traditional tools of tort law. They apply longstanding doctrines passed down and modified through the evolution of the common law to isolate and grapple with the issues that the long history of human wrongdoing tells us are relevant to liability: How much did the gun show promoter know about the illegal sales? Did it do anything to promote them? How foreseeable was the wrongful conduct? And were the gun show organizer's policies and procedures reasonable under the circumstances?

For example, in *Pavlides v. Niles Gun Show, Inc.*,[51] the Court of Appeals of Ohio considered the viability of a negligence claim against the promoter of a gun show from which firearms were stolen and later used to commit a crime.[52] In that case, the gun show promoter permitted a group of four teenagers—ages thirteen to seventeen—to enter a gun show unaccompanied by an adult.[53] The group proceeded to steal numerous firearms from the third-party vendors at the show, including a .38 caliber handgun.[54] One of the teens later explained that the firearms were "'just laying around' on tables," that they "just pick[ed] them up and walk[ed] away with them," and that "it was easy."[55] Finding ammunition was also easy. The thirteen-year-old purchased matching .38 caliber ammunition from a vendor at the show unaccompanied by an adult and without proof of age.[56] The next day, the teens stole a Chevrolet Camaro and took it for a joy ride in the snow, intentionally swerving and sliding the car into trash cans as they went.[57] When they were stopped by two officers in a police car, one of the teens pulled out the .38 caliber handgun and shot the first officer in the chest.[58] As the second officer turned and ran, the teen shot him in the back of the head.[59]

---

[51] 637 N.E.2d 404 (Ohio Ct. App. 1994).
[52] *Id.* at 406–07.
[53] *Id.*
[54] *Id.*
[55] *Id.* at 407 (alterations in original) (quoting Edward A. Tilley III).
[56] *Id.*
[57] *Id.*

The facts of the case are tragic. A gang of unsupervised, unruly teens got into trouble, made extraordinarily poor choices, and the situation turned dire. Ordinarily, of course, tort law would not recognize a claim against a gun show promoter based on the later misuse of firearms stolen from its events. The gun show promoter was separated from the later tortious conduct by multiple actors—third-party vendors and teens—and multiple instances of criminal conduct.[60] But ordinarily is not always. Sometimes liability is warranted even where intervening third-party conduct contributes to the harm. Discerning the difference requires application in each case of tort law's nuanced rules governing liability for third-party harms.[61]

Ultimately, the *Pavlides* Court determined that the plaintiff police officers pleaded a cognizable claim against the gun show promoter.[62] Considering the foreseeability of the teens' actions and the gun show promoter's duty to the third-party police officers, the court reasoned that "reasonable minds certainly could conclude that unsecured firearms present an attractive if not irresistible lure to children."[63] The court concluded that the police officer shootings were a foreseeable result of both the promoter's lax admission procedures, which failed to identify the teens as unaccompanied minors, and its failure to require that vendors secure their weapons or refuse sales to minors, especially given the promoter's knowledge that weapons had been stolen from past shows.[64] As for proximate cause, the court reasoned that although proximate cause is ordinarily lacking where unforeseeable criminal acts interrupt the chain of causation, unsupervised teens and firearms are a dangerous combination and a jury could find the criminal acts to be a foreseeable result of the promoter's lack of safeguards.[65] Slight factual variations or differences in governing law could easily produce different results.[66] The analysis requires fact-specific application of doctrines crafted to address difficult questions of causation, duty, and liability in cases involving multiple causal actors and intervening wrongful conduct.

---

58 *Id.*

59 *Id.*

60 *See* Restatement (Third) of Torts: Liab. for Physical & Emotional Harm §§ 19, 34 (Am. L. Inst. 2010) (explaining that where an intervening third-party act contributes as a factual cause of harm, an actor's liability is limited to harms foreseeably flowing from her lack of care, but that third-party wrongdoing will sometimes be a foreseeable risk that she must take reasonable care to avoid).

61 *See id.* § 34 cmt. a (describing the "extensive rules" that have historically governed "when intervening acts become sufficient to 'supersede' an actor's earlier tortious conduct" but also noting that tort law's evolution in recent decades toward liability and damage awards based on comparative fault has reduced the need for proximate and superseding cause rules that bar liability altogether).

62 *See Pavlides*, 637 N.E.2d at 409–10.

63 *Id.* at 409.

64 *Id.*

### 2. The Virtual World

In the virtual world though, the analysis is completely different. With Section 230 of the CDA, Congress preempted most claims against websites when those claims involve actions by third parties, such as their users.[67] Because the modern internet's major players are now built around user-created data—for example, Facebook, Twitter, YouTube, and Snapchat—Section 230 has had vast consequences for legal actions against internet entities. The validity of claims brought against such entities for third-party harms is determined not by the intricate legal doctrines designed for the purpose, but by courts' attempts to parse a single phrase from Congress: "No provider or user of an interactive computer service shall be treated as the publisher or speaker of any information provided by" someone else.[68]

When Yasmeen Daniel brought her virtual-world claim against Armslist.com for its role in the shooting death of her mother, the court dismissed the claim under Section 230 on immunity grounds without reaching the cognizability of her claim.[69] It did not consider the foreseeability and care related questions that are so crucial to the analysis in the strikingly similar, physical-world *Pavlides* decision.[70] The *Armslist* court never considered the adequacy of Armslist.com's precautionary measures at all. Armslist.com would have been entitled to dismissal under Section 230 even if its conduct had exceeded the negligence in *Pavlides* and rose to the level of the deliberate recklessness of the hypothetical Illegal Gun Sales Warehouse. Rather than questions about knowledge, foreseeability, and fault, the viability of claims in the virtual world often turns on courts' attempts to parse Section 230 to determine whether a defendant was an "information content provider" of the material on its website or whether it was a mere conduit, a "computer service" providing access to that information.[71]

---

65 *See id.* at 409–10.

66 *See, e.g.*, Bloxham v. Glock Inc., 53 P.3d 196, 198–200 (Ariz. Ct. App. 2002) (applying Arizona's test for duty to third parties and concluding on similar facts that gun show promoter had no duty to third-party victims regardless of foreseeability); Hamilton v. Beretta U.S.A. Corp., 750 N.E.2d 1055, 1060–62 (N.Y. 2001) (opinion of Wesley, J.) (concluding handgun manufacturers do not owe victims a duty to exercise care in marketing and distribution); Valentine v. On Target, Inc., 727 A.2d 947, 952 (Md. 1999) (finding gun dealer had no duty to third party victim of gun stolen from dealer's store where there was no lack of due care storing the weapon, no knowledge of circumstances that would increase the probability of theft, and no special relationship between dealer and victim); Estate of Strever v. Cline, 924 P.2d 666, 674, 668–71 (Mont. 1996) (concluding that owner of unlocked pickup truck with pistol and ammunition stored under the seat had a duty to three teenagers who stole the pistol and ammunition—which one of the teens later inadvertently fired, shooting and killing his friend—but the truck owner's breach of that duty was not a proximate cause of the death).

67 *See* 47 U.S.C. § 230(c)(2).

68 *Id.* § 230(c)(1).

69 *See* Daniel v. Armslist, LLC, 926 N.W.2d 710, 727 (Wis. 2019), *cert. denied*, 140 S. Ct. 562 (2019).

As discussed in more detail below,[72] this approach is problematic for several reasons. First, Section 230 is undertextured[73] in that it is insufficiently precise to address the hard cases. Second, the law was designed for a different purpose; is poorly suited to resolving questions of when to impose liability for third-party harms; and, unlike the common law, has not evolved to keep pace with shifting technologies. Third, even assuming Congress intended to create a separate set of rules to govern internet tort claims for third-party harms, that dichotomous framework does not fit the modern internet. It is increasingly problematic to treat online and offline defendants differently in a world where so much economic and social life now takes place in the virtual sphere.

### C. The Birth of the Disparity

#### 1. Section 230's Text and Origin

Congress enacted Section 230 of the CDA to promote "decency" on the internet and address the problem of children's unrestricted access to pornography and other offensive material on the internet by allowing online entities to censor such content without fear of being held liable for whatever content they failed to censor.[74] The law pro-

---

70    *See Pavlides*, 637 N.E.2d at 409–10.

71    47 U.S.C. § 230(c).

72    *See infra* Part II.

73    *See* H.L.A. Hart, The Concept of Law 124–36 (3d ed. 2012) (discussing statutory law's "open texture" and the idea that a statute's treatment of some areas of conduct must be left for development by courts).

74    *See* 47 U.S.C. § 230(b) ("It is the policy of the United States . . . to remove disincentives for the development and utilization of blocking and filtering technologies . . . ."); H.R. Rep. No. 104-458, at 194 (1996) (noting the "important federal policy of empowering parents to determine the content of communications their children receive"); 141 Cong. Rec. S8,088 (daily ed. June 9, moted online freedom of expression and the then-burgeoning, and potentially fragile, array of communication and informational resources that were becoming available on the internet by guaranteeing online entities' ability to relay and host the massive volume of content flowing into their systems without incurring liability for its contents.[75] With these aims in mind, Congress crafted Section 230 to read, in relevant part, as follows:

> (c) Protection for "Good Samaritan" blocking and screening of offensive material
>> (1) Treatment of publisher or speaker
>>> No provider or user of an interactive computer service shall be treated as the publisher or speaker of any information provided by another information content provider.
>>
>> (2) Civil liability
>>> No provider or user of an interactive computer service shall be held liable on account of—
>>> (A) any action voluntarily taken in good faith to restrict access to or availability of material that the provider or user considers to be obscene, lewd, lascivious, filthy, excessively violent, harassing, or otherwise objectionable, whether or not such material is constitutionally protected . . .
>
> (f) Definitions . . .
>> (3) Information content provider
>>> The term "information content provider" means any person or entity that is responsible, in whole or in part, for the creation or development of information provided through the Internet or any other interactive computer service.[76]

The section includes two key provisions. First, Section 230(c)(1) immunizes "interactive computer service[s]"—websites, apps, internet

---

1995) (statement of Sen. J. James Exon) ("[T]he worst, most vile, most perverse pornography is only a few click-click-clicks away from any child on the Internet."); *see also* Zeran v. Am. Online, Inc., 129 F.3d 327, 331 (4th Cir. 1997) ("Another important purpose of § 230 was to encourage service providers to self-regulate the dissemination of offensive material over their services.").

[75] *See* 47 U.S.C. § 230(a)(3) ("The Internet and other interactive computer services offer a forum for a true diversity of political discourse, unique opportunities for cultural development, and myriad avenues for intellectual activity."); *Zeran*, 129 F.3d at 330 ("Congress recognized the threat that tort-based lawsuits pose to freedom of speech in the new and burgeoning Internet medium.").

[76] 47 U.S.C. § 230(c), (f)(3).

service providers, and the like—from any theory of liability that depends upon their being "treated as the publisher or speaker of any information provided by" someone else.[77] As a result, the Verizons, Twitters, and Facebooks of the world can provide access to the internet and to the posts of their users without fear, for example, of incurring tort liability when a user's tweet or post defames a third party. Any defamation claim against Facebook for a user's post would require "treat[ing it] as the publisher or speaker"[78] of that post and would thus be barred by Section 230.

Second, pursuant to Section 230(c)(2), "[n]o provider . . . of an interactive computer service shall be held liable on account of . . . any action voluntarily taken in good faith to restrict access to" objectionable content.[79] Congress believed that the prospect of potential tort liability was limiting service providers' willingness to deploy content-filtering technologies to help parents limit their children's access to objectionable material and that this provision would broaden the availability of such technologies.[80]

Rather than directly regulating entities and requiring them to censor content, Section 230 operates by removing a major disincentive to censorship: the threat of tort liability. In crafting Section 230, Congress had firmly in mind the then-recent decision *Stratton Oakmont, Inc. v. Prodigy Services Co.*,[81] in which a New York state trial court held an internet service provider liable for defamatory content posted by a third party on one of the service's message boards.[82] *Stratton Oakmont* had reasoned that because Prodigy took steps to review and screen offensive content, it had taken on the role of a newspaper-like publisher rather than a mere distributor and could therefore be held liable for repeating the defamer's words.[83] That is, because Prodigy screened and "restrict[ed] access to"[84] some objectionable content, it was deemed to have ratified[85] all content that it did not screen and

---

   77   *Id.* § 230(c).
   78   *Id.*
   79   *Id.* § 230(c)(2).
   80   *See id.* § 230(b)(3)–(4); H.R. Rep. No. 104-458, at 194 (1996).
   81   No. 031063/94, 1995 WL 323710 (N.Y. Sup. Ct. May 24, 1995), *superseded by statute*, Communications Decency Act ("CDA") of 1996, Pub. L. No. 104-104, tit. V, 110 Stat. 56, 133–43 (codified at 47 U.S.C. § 230).
   82   *Id.* at *5; H.R. Rep. No. 104-458, at 194 ("One of the specific purposes of this section is to overrule *Stratton-Oakmont v. Prodigy* . . . .").
   83   *Stratton Oakmont*, 1995 WL 323710, at *5.
   84   47 U.S.C. § 230(c)(2)(A).
   85   Although applied by the court in *Stratton Oakmont* to a new context, the tort theory of vicarious liability for ratification of tortious speech by failure to remove it long predates the could be "treated as the publisher or speaker"[86] of that content for purposes of the plaintiff's defamation claim. With Section 230, Congress emphatically rejected that theory of liability, freeing websites, internet service providers, and others to screen objectionable or simply off-topic, third-party-created content without fear of being held liable as the publisher or speaker[87] of whatever content they choose not to censor.[88]

### 2. The Internet-Fueling Liability Shield

By almost every account, Section 230 has been an astounding success. The federalization of internet law and elimination of any obligation on online entities to moderate content helped to fuel the internet's explosive growth from the 1990s through the 2000s. As Jeff Kosseff has observed, "[t]he early Section 230 court opinions could not have come at a better time . . . [a]s the industry was slowly recovering from the dot-com implosion of 2000 and 2001," and companies were just beginning to develop a new generation of two-way websites centered around user content—sites like YouTube, Facebook, Wikipedia, and Yelp, that have come to dominate the modern internet.[89] Such sites "provided once-voiceless consumers with a megaphone to warn others against companies' scams" and "revolutionized how Americans receive information."[90]

Section 230 has supported the internet's development in three important ways. First, Section 230 shields online entities from an economically crippling duty to moderate the content flowing through their systems. Think of online platforms like Facebook or Twitter, for example. Their users create and upload an almost inconceivable volume of content each day. Requiring such entities to review that con-

---

internet. *See* Gregory M. Dickinson, Note, *An Interpretive Framework for Narrower Immunity Under Section 230 of the Communications Decency Act*, 33 HARV. J.L. & PUB. POL'Y 863, 877–78 (2010); *see also* RESTATEMENT (SECOND) OF TORTS § 577(2) (AM. L. INST. 1977) ("One who intentionally and unreasonably fails to remove defamatory matter that he knows to be exhibited on land or chattels in his possession or under his control is subject to liability for its continued publication."); Tacket v. Gen. Motors Corp., 836 F.2d 1042, 1046 (7th Cir. 1987) (opinion of Easterbrook, J.) (discussing theory of liability through adoption of another's publication where employer failed to remove an allegedly libelous sign from factory wall).

86  47 U.S.C. § 230(c)(1).

87  *See id.* § 230(c).

88  For an insightful discussion of Section 230's interaction with pre-internet defamation law and affirmative duties to intervene, see JOHN C.P. GOLDBERG & BENJAMIN C. ZIPURSKY, RECOGNIZING WRONGS 319–39 (2020).

89  KOSSEFF, *supra* note 12, at 120–21.

90  *Id.*

tent and remove unlawful material would impose an insurmountable logistical and financial burden on internet entities and undermine the internet as we know it.[91] By immunizing online entities against lawsuits related to third-party content, Section 230 ensures that the costs of moderating user-created content do not stifle the growth of internet platforms.

Second, and relatedly, Section 230 protects against collateral censorship of users' speech. Were online entities at risk of legal liability whenever one of their users posted something unlawful, platforms might decide to block their users from posting even slightly risky material to avoid the cost of moderating content and the risk of legal liability for failing to do so.[92] Why risk posting content that could subject the company to liability when the platform, compared to the speaker herself, has no intrinsic interest in publicizing the message? Most likely to engage in widespread censorship would be platforms like Twitter, which transmit so much content that they could not possibly hope to screen it all. But the incentive to censor would press even low-volume sites, like individual blogs, whose operators typically lack the resources of larger entities to respond to problematic material.[93] Without Section 230's protections, the legal risks of hosting user-created content would push entities of all sizes to dramatically alter their current practices. Platforms might require that user's posts be manually prescreened before becoming visible to the public; they might adopt automated censorship tools calibrated to let through only the most benign speech; or they might decide to eliminate user-created content from their sites altogether by removing comment posting functionality. The result would be the elimination of decades of free speech advances built on inexpensive and free flowing internet publishing technologies.[94]

---

[91] *See* Mark A. Lemley, *Rationalizing Internet Safe Harbors*, 6 J. ON TELECOMMS. & HIGH TECH. L. 101, 101–02 (2007) (noting the billions of web pages indexed by Google's search engine and observing that if it or other "Internet intermediaries were liable every time someone posted problematic content on the Internet, the resulting threat of liability and effort at rights clearance would debilitate the Internet"); *see also* Brief of Amicus Curiae Electronic Frontier Foundation at 3–4, Daniel v. Armslist, LLC, 926 N.W.2d 710, 727 (Wis. 2019) (No. 2017-AP-344) (arguing that platforms are unable "both logistically and financially" to oversee the content on their platforms "given the incredible volume of content generated by platform users"), *cert. denied*, 140 S. Ct. 562 (2019).

[92] Wu, *supra* note 6, at 298–301.

[93] *See id.* at 301 (reasoning that the collateral censorship problem extends to both low- and high-volume intermediaries and that some of the highest value speech, like corporate whistleblowing, is risky for intermediaries and "the most likely to be censored").

[94] *See id.* at 298–99 (observing that pre-internet speech was "limited to those who were

These two concerns—protecting online entities from the burden of content moderation and protecting users from resultant collateral censorship—have roots that long predate Section 230. The first courts to assess whether and when online entities should be liable for their content sought guidance from the nearest physical-world analogue—tort law governing when hard-copy book and newspaper publishers or the bookstores and news vendors distributing their publications should be liable for defamatory content they contain.[95]

In its early days, the internet more closely resembled the traditional publication world, and pre–Section 230 courts quickly adopted publication-world principles. In particular, courts noted that print media distributors generally are not liable for defamatory content within their pages because requiring them to scour the pages of books and newspapers they sell would be an impossible burden.[96] Early courts noted the similar deluge of content that passed through internet service providers like CompuServe and America Online and reasoned that they too should ordinarily not be liable for the content they make available because, like a "public library, book store, or newsstand," it would not be feasible for the online entity "to examine every publication . . . for potentially defamatory statements"[97] and "[s]uch a rule would be an impermissible burden on the First Amendment."[98] Congress made that principle part of federal law by enacting Section 230. Section 230 fosters free expression on the internet[99] by federalizing the principle that online entities, like print distributors,[100] should not

---

able to get past the old gatekeepers—newspapers, book publishers, retailers, and the like" but that "[n]ow all that is needed is an Internet connection"); *see also* Eugene Volokh, *Cheap Speech and What It Will Do*, 104 YALE L.J. 1805, 1806–07 (1995) (noting that historically the right to free speech has favored popular or well-funded ideas, but predicting, presciently, that new information technologies would dramatically reduce the costs of distributing speech and create a more diverse and democratic environment).

[95] *See* KOSSEFF, *supra* note 12, at 39–40.

[96] *See* Brent Skorup & Jennifer Huddleston, *The Erosion of Publisher Liability in American Law, Section 230, and the Future of Online Curation*, 72 OKLA. L. REV. 635, 638–46 (2020) (detailing evolution of defamation liability for publishers and distributors in the twentieth century from strict liability to a fault-based regime that required knowledge or recklessness because of the impossible burden of moderating mass media).

[97] Cubby, Inc. v. CompuServe Inc., 776 F. Supp. 135, 140 (S.D.N.Y. 1991).

[98] *Id.* (quoting Lerman v. Flynt Distrib. Co., 745 F.2d 123, 139 (2d Cir. 1984)).

[99] Section 230's other, often overlooked, feature was to ensure that even if entities voluntarily decided to filter content, they would not thereby incur potential liability. *See* 47 U.S.C. § 230(c)(2) (immunizing "Good Samaritan" entities from liability predicated on content-filtering efforts).

[100] Section 230 grants online entities even broader immunity than did the common law, in that it immunizes even entities who have actual or constructive knowledge of unlawful content on their platforms. *See id.* § 230(c).

be liable for third-party-created content available through their services.[101]

Third, Section 230 offers several procedural advantages in litigation that reduce the costs of defending lawsuits and allow online entities to commit more of their resources to developing and managing products. Section 230's primary advantages over alternative defenses stem from its simplicity: If an online entity did not author the content at issue, it cannot be liable for any harms flowing from that content.[102] The defense provides uniform protection against any cause of action,[103] which prevents plaintiffs from circumventing the defense through creative pleading.[104] Because the defendant's scienter is irrelevant and the defense relies on facts often obvious on the face of the complaint (for example, who authored the relevant content), the defendants can often successfully assert the defense early, on the pleadings alone, and thereby avoid discovery and minimize litigation costs.[105] And because the defense is grounded in federal, rather than state, law it applies in every jurisdiction, improving national uniformity and reducing costly state-by-state compliance analysis.[106]

---

101 *See* Zeran v. Am. Online, Inc., 129 F.3d 327, 331 (4th Cir. 1997) (interpreting Section 230 and reasoning that "[i]t would be impossible for service providers to screen each of their millions of postings," that providers might respond by restricting user content, and that "Congress considered the weight of the speech interests implicated and chose to immunize service providers to avoid any such restrictive effect").

102 *See* 47 U.S.C. § 230. Eric Goldman discusses Section 230's procedural advantages in his recent essay contrasting Section 230's protections with those of the First Amendment. *See* Goldman, *supra* note 6, at 39–44. Many of his observations apply outside the First Amendment context as well. *See id.*

103 *See* Barnes v. Yahoo!, Inc., 570 F.3d 1096, 1101–02 (9th Cir. 2009) ("[W]hat matters is not the name of the cause of action—defamation versus negligence versus intentional infliction of emotional distress—what matters is whether the cause of action inherently requires the court to treat the defendant as the 'publisher or speaker' of content provided by another.").

104 Some courts have limited immunity to causes of action for which publication is a required element. *See, e.g.*, City of Chicago v. StubHub!, Inc., 624 F.3d 363, 366 (7th Cir. 2010) (reasoning that Section 230 "limits who may be called the publisher [or speaker] of information that appears online," which "might matter to liability for defamation, obscenity, or copyright infringement" but not "Chicago's amusement tax"). Given courts' broad conception of what editorial functions Section 230 protects, however, this requirement is almost always satisfied. *See* Doe v. Backpage.com, LLC, 817 F.3d 12, 19 (1st Cir. 2016) (noting that "[t]he broad construction accorded to section 230 as a whole has resulted in a capacious conception of what it means to treat a website operator as [a] publisher or speaker" and that section 230 has accordingly been applied to "a wide variety of causes of action, including housing discrimination, negligence, and securities fraud and cyberstalking" (citations omitted)).

105 *See* Engine, Section 230: Cost Report 2 (2019), https://static1.squarespace.com/static/571681753c44d835a440c8b5/t/5c8168cae5e5f04b9a30e84e/1551984843007/Engine_Primer_230cost2019.pdf [https://perma.cc/NC52-XGBP] (concluding, based on interviews with defense attorneys regarding litigation costs, that Section 230 would be far less effective were it "more difficult for startups to rely on it early in a lawsuit, before litigation costs escalate to the point where settling is less expensive that actually winning").

### 3. A Tripartite Statute in a Multipartite World

Despite its continuing importance to the online world, Section 230 now shows its age. The 1996 statute is designed for an internet that functions to distribute third-party content. Like print and broadcast media distributors before them, online entities often serve as access points for vast stores of third-party-created content—so vast that they could never hope to police their content.[107] Section 230 adopts tort law's treatment of mass-media distributors and grants online entities broad federal immunity from claims related to third-party content.[108]

To do so, Section 230 divides the online world into three camps:[109] (1) "information content provider[s],"[110] analogous to traditional publishers, who create informational content for internet readers; (2) providers of "interactive computer service[s],"[111] such as CompuServe, Prodigy, and, today, Verizon and Comcast, that provide users with the internet access necessary to read that informational content; and (3) users of "interactive computer service[s],"[112] the internet users who purchase internet access from an internet service provider and use that access to consume informational content made by online content creators.[113]

In such a neatly divisible virtual world, potential wrongdoers are easy to categorize. Any wrongdoing must be attributable to its active participants—the internet content creators who author content—not the passive internet service providers and their users who merely pro-vide access to and view that content. The solution seemed obvious and uncontroversial when Congress set about to undo the New York state trial court's questionable decision in *Stratton Oakmont*, which had permitted a claim against an internet service provider for defamatory content posted by a third party.[114] Congress immunized the internet's passive participants from liability with Section 230, by mandating that none but an "information content provider" be held accountable for internet content.[115]

---

[106] Goldman, *supra* note 6, at 44.

[107] *See* Skorup & Huddleston, *supra* note 96, at 638–46 (describing defamation law's evolution from strict liability to a fault-based regime in the print media context given the difficulty that republishers would face trying to moderate content); Lemley, *supra* note 91, at 101–02 (noting the same problem in the online context).

[108] *See* Skorup & Huddleston, *supra* note 96, at 649.

[109] Both the congressional findings supporting the enactment of Section 230 and its operative provisions adopt a tripartite view of the internet. The internet, for example, is described as one among "other interactive computer services" that serve as communications channels for "political discourse" and "intellectual activity," 47 U.S.C. § 230(a)(3), and for transmission to and receipt by users of "information." *Id.* § 230(a)(2).

[110] *Id.* § 230(f)(3).

[111] *Id.* § 230(f)(2).

[112] *Id.* § 230(c)(1).

[113] For a more detailed discussion of the contrast between the closed communities and curated content of the early internet and the interactive websites of today, see KOSSEFF, *supra* note 12, at 36–38, 177–80.

The problem is that the internet was never entirely publication-centric or so neatly divisible. Today, it is even less so than it was in 1996. While Section 230 was being debated in Congress, e-commerce was starting to take root on the internet. Amazon.com sold its first book in July 1995,[116] and AuctionWeb, later to become eBay, made its first sale just a few months later, in September 1995.[117] These and other online entities quickly began to strain Section 230's tripartite framework, which is designed for a world comprised of content authors, transmitters, and readers—not for the virtual world of e-commerce and online services.[118] Categorizing all internet entities as either content authors, and potentially liable, or as nonauthors, and therefore immune, makes little sense when virtual world wrongdoing often involves no content creation at all. The dissonance between the virtual world and Section 230's tripartite, publishing-world structure has grown in the decades following the statute's enactment.[119]

---

[114] *See* Stratton Oakmont, Inc. v. Prodigy Servs. Co., No. 031063/94, 1995 WL 323710 (N.Y. Sup. Ct. May 24, 1995), *superseded by statute*, Communications Decency Act ("CDA") of 1996, Pub. L. No. 104-104, tit. V, 110 Stat. 56, 133–43 (codified at 47 U.S.C. § 230).

[115] Early courts then cemented this division by applying Section 230 broadly to bar all claims other than those against "content providers," even where those claims are only tangentially related to any internet content. *See supra* Section I.B.

[116] *See* Mark Hall, *Amazon.com*, Britannica (Apr. 9, 2020), https://www.britannica.com/topic/Amazoncom [https://perma.cc/ZHB4-EW58].

[117] *See Our History*, eBay, https://www.ebayinc.com/company/our-history/ [https://perma.cc/7LVN-3TUG].

[118] For a discussion and critique of a parallel issue in First Amendment law, which treats the electronic data that drive code-dependent technologies as expressive speech, see Kyle Langvardt, *The Doctrinal Toll of "Information as Speech*," 47 Loy. U. Chi. L.J. 761 (2016).

[119] European nations have faced a similar problem. The European Union's E-Commerce Directive currently immunizes "[m]ere conduits," "[c]aching," and "[h]osting" entities from liability. Council Directive 2000/31, 2000 O.J. (L 178) 1, 12–14 (EC). A recent report commissioned by the European Parliament noted that these categories "are incredibly contested," "that very few digital operators truly fit these categories," and that under the regulation a "gap has flourished that has the dual effect of stifling innovation and is inadequate for providing suitable safeguards." Melanie Smith, Eur. Parliament, Enforcement and Cooperation Between Member States: E-Commerce and the Future Digital Services Act 11 (2020).

One major change has been the internet's dramatic shift toward specialization, interactivity, and the offering of real-world goods and services. When Section 230 was enacted in 1996, less than eight percent of Americans had access to the internet, and for almost all of them, access was through painfully slow dial-up connections.[120] Not only were connections sluggish, but there was also little to do or see when connected. Relatively few entities had websites, and the internet was small enough[121] that search engines could still rely on human discovery and classification of pages rather than automated web crawlers.[122] Google and even webmail were still years away.[123] With many important interactive web development technologies yet to be invented or still in their infancy,[124] websites tended to be merely static repositories from which users could retrieve informational content. A bank's website, for example, might include a page displaying its hours of operation, but it would have included no online banking functionality. Online banking was virtually unheard of, and almost all banking was still conducted at brick-and-mortar locations.[125] With so few Americans online and so few options for those who were, the internet was used infrequently and consisted largely of websites targeted toward general audiences and acting as content-communication mecha-

---

[120] *See* Farhad Manjoo, *Jurassic Web*, SLATE (Feb. 24, 2009, 5:33 PM), https://slate.com/technology/2009/02/the-unrecognizable-internet-of-1996.html [https://perma.cc/M3EC-BLQB].

[121] When Congress took up consideration of the CDA in 1995, the internet contained somewhere between 25,000 and 250,000 websites. *See Total Number of Websites*, INTERNET LIVE STATS, https://www.internetlivestats.com/total-number-of-websites/#trend [https://perma.cc/T75Y-KCYM]. In the years since, the number of websites has risen at a staggering rate; the internet now contains a barely comprehensible 1.7 billion websites. *See id.*

[122] *See* Manjoo, *supra* note 120.

[123] *See From the Garage to the Googleplex*, GOOGLE, https://about.google/our-story [https://perma.cc/UPL7-BJLG] (discussing the timing of Google's founding).

[124] For example, JavaScript, the primary driver of interactive features such as text boxes, buttons, drop-down lists, etc. on most websites today, was still being created when Congress enacted Section 230. *See* Gabriel Lebec, *JavaScript: A History for Beginners*, COURSE REP. (Mar. 12, 2019), https://www.coursereport.com/blog/history-of-javascript [https://perma.cc/6DDU-VDYX]. Similarly, interactive video media did not become mainstream until the release of Adobe Flash 5 in 2000. *See* Jay Hoffmann, *Flash and Its History on the Web*, HIST. OF THE WEB (Aug. 7, 2017), https://thehistoryoftheweb.com/the-story-of-flash [https://perma.cc/5448-BSGH]. Additionally, Adobe Flash's successor, HTML5, was not finalized by the World Wide Web Consortium until 2014. *See* W3C, HTML5: RECOMMENDATION (2014), https://www.w3.org/TR/2014/REC-html5-20141028 [https://perma.cc/625H-Z7WF].

[125] Stanford Federal Credit Union was the first U.S. financial institution to offer internet banking in 1994. *See* Laura Woods, *How Online Banking Evolved into a Mainstream Financial Tool*, MOTLEY FOOL (Nov. 9, 2014, 9:00 AM), https://www.fool.com/investing/general/2014/11/09/how-online-banking-evolved-into-a-mainstream-finan.aspx [https://perma.cc/2C2G-KDV8]. Online banking did not become common, however, until the early 2000s. *See id.*

nisms. Even those Americans who had internet access spent fewer than thirty minutes per month online.[126]

Contrast that early static and content-driven internet with the interactive platforms and web services of today. In the decades since 1996, the percentage of Americans with internet access has risen tenfold, to almost ninety percent.[127] Internet use worldwide has followed the same trend, to the point where by 2019, almost sixty percent of the world's 7.7 billion people were connected to the internet.[128] As more and more internet users have come online, websites have become increasingly specialized. Meanwhile, advances in interactive web development and backend data storage and processing technologies have enabled a shift from the static information-repository-style websites common in the 1990s to the dynamic and interactive web services that users are familiar with today. Static content and information retrieval remain important,[129] but today's internet is much more than that.

Go to your bank's website today and you may still find a page displaying the hours during which you can visit the bank's physical locations.[130] But there is almost never a need to do so because internet users can now conduct their banking transactions online. And banking is no aberration. Nearly every aspect of human life has been reproduced in the virtual world, which now coexists in parallel with the physical world from which it sprung. Shopping for a young nephew's birthday gift? Walking through Toys "R" Us[131] might now be clicking

---

[126] *See* Manjoo, *supra* note 120.

[127] *Internet/Broadband Fact Sheet*, Pew Rsch. Ctr. (June 12, 2019), https://www.pewresearch.org/internet/fact-sheet/internet-broadband [https://perma.cc/5C5X-LXS7].

[128] J. Clement, *Worldwide Digital Population as of July 2020*, Statista (Oct. 29, 2020), https://statista.com/statistics/617136/digital-population-worldwide/ [https://perma.cc/M3DQ-L6SW].

[129] The online encyclopedia, Wikipedia, is an especially prominent example. Despite hosting primarily static, informational content, it ranks as one of the internet's twenty most frequently visited websites. *See The Top 500 Sites on the Web*, Alexa, https://www.alexa.com/topsites [https://perma.cc/P7BW-9MWC]. Even a site like Wikipedia, however, is significantly more interactive than those of the early internet. Wikipedia's search and community-editing functionality would not have been possible when Section 230 was enacted. *See* Ryan Dube, *The Origins of Wikipedia: How It Came to Be [Geek History Lesson]*, Make Use Of (July 5, 2012), https://www.makeuseof.com/tag/origins-wikipedia-si-title/ [https://perma.cc/6Y3F-S92Q] (explaining that Wikipedia, with its new collaborative platform, arose in 2000, several years after the enactment of Section 230).

[130] *See, e.g.*, *550 Fifth Avenue*, Bank of Am., https://locators.bankofamerica.com/ny/newyork/financial-centers-new-york-16767.html [https://perma.cc/6X5X-NHZ4].

[131] After declaring bankruptcy in 2017, Toys "R" Us closed all its U.S. stores. Abha Bhattarai, *Toys R Us Is Back from the Dead, but Its New Stores Are Unrecognizable*, Wash. Post (Jul. 18, 2019, 7:30 AM MDT), https://www.washingtonpost.com/business/2019/07/18/toys-r-us-is-back-dead-its-new-stores-are-unrecognizable [https://perma.cc/TG42-FQ3Z]. It has since re-

through Amazon.com.[132] Buying a new house? No problem. Try Zillow.[133] Considering a garage sale to get rid of your old junk? In the virtual world, try VarageSale.[134] Want to check your kid's grades or argue with her teachers? Your school district probably has a website for that too.[135] Of course, the phenomenon is not limited to noble pursuits like helicopter parenting. Internet destinations have also cropped up for users to make sex-for-pay arrangements,[136] buy or sell illicit drugs,[137] or share revenge porn photos or videos of former sexual partners.[138]

---

opened two locations, with a focus on "open play areas, interactive displays and spaces for special events and birthday parties." *Id.*

[132] Sales on Amazon.com account for almost half of all online retail sales in the United States and four to five percent of total retail sales. *See* Scott Shane, *Prime Mover: How Amazon Wove Itself into the Life of an American City*, N.Y. Times (Nov. 30, 2019) https://www.nytimes.com/2019/11/30/business/amazon-baltimore.html [https://perma.cc/JY5S-FNQ6].

[133] Zillow operates a database of more than 110 million U.S. homes and describes itself as "the leading real estate and rental marketplace." *See About*, Zillow, https://www.zillow.com/corp/About.htm [https://perma.cc/GZ6A-UBBW].

[134] VarageSale is a marketplace for selling second-hand goods. *See What is VarageSale?*, VarageSale (Dec. 8, 2020), https://help.varagesale.com/article/161-what-is-varagesale [https://perma.cc/59ZS-F6P4]. It describes itself as "the safer, smoother way to buy and sell locally." *Id.* It differs from more well-known sites like eBay by focusing on local, in-person sales and from sites like Craigslist by providing a highly structured sale and payment platform. *See id.*

[135] Many school districts now rely on cloud-based software for tracking student progress and attendance and reporting grades to students and parents. *See School Administration Software*, Capterra, https://www.capterra.com/school-administration-software [https://perma.cc/VXB6-Z6VP] (ranking and reviewing various school administration tools).

[136] Such websites, which include Slixa, SeekingArrangement, and the now defunct Backpage, are often thinly disguised as forums for dating or unspecified "adult services." *See* Nicholas Kristof, *'Every Parent's Nightmare,'* N.Y. Times, Mar. 11, 2016, at 9; Rachel Weiner, *Man Sentenced to 15 Years for Sex with Teen Met Online*, Wash. Post, Jan. 21, 2018, at C6; Tom Jackman & Mark Berman, *Authorities Take Down Backpage.com*, Wash. Post, Apr. 8, 2018, at A7. Even websites and apps intended for traditional dating can be misused by sexual predators to connect with prospective victims. *See* Irina D. Manta, *Tinder Lies*, 54 Wake Forest L. Rev. 207 (2019).

[137] Since the 2013 closure of Silk Road, an online black market hidden on the dark web, other markets have sprung up to take its place. *See* Nathaniel Popper, *Dark Web Drug Sellers Dodge Police Crackdowns*, N.Y. Times (June 11, 2019), https://www.nytimes.com/2019/06/11/technology/online-dark-web-drug-markets.html [https://perma.cc/93LJ-7JLQ].

[138] *See* Deanna Paul, *Courts Wrestle with Free Speech vs. Revenge Porn*, Wash. Post, Dec. 27, 2019, at A6; Carrie Goldberg, *How Google Has Destroyed the Lives of Revenge Porn Victims*, N.Y. Post (Aug. 17, 2019, 1:10 PM), https://nypost.com/2019/08/17/how-google-has-destroyed-the-lives-of-revenge-porn-victims [https://perma.cc/DR3A-MEHC]; People v. Austin, 155 N.E.3d 439, 471 (Ill. 2019) (upholding Illinois revenge porn law as not improperly restricting freedom of speech), *cert. denied*, 141 S. Ct. 233 (2020). *See generally* Danielle Keats Citron & Mary Anne Franks, *Criminalizing Revenge Porn*, 49 Wake Forest L. Rev. 345 (2014) (describing the problem of nonconsensual distribution of private sexual images and arguing for the criminalization of revenge porn).

In short, technological advances and a dramatic rise in internet users have completely transformed the internet landscape into a complete virtual world that is far too complex to be governed by Section 230's tripartite, publication-centric model.

## II. Section 230 Is Poorly Tailored

Despite these dramatic changes to the internet landscape, online entities' liability[139] for third-party conduct is still governed by a bright line rule designed for the internet of 1996: Content creators may be sued, but online entities that use that content are immune.[140] Convenient and apt as the rule may be for publication-related entities, universal application of the rule to today's internet results in unsettling disparities between online and offline entities.[141] More specifically, Section 230's content authorship-based test for immunity is problematic in cases where a plaintiff's injury is causally connected to third-party-created content, but the defendant's alleged wrongdoing is not based on a failure to moderate that content. This Part analyzes three categories of cases in which Section 230 is a poor fit for the modern internet. That analysis sets the stage to consider, in Part III, how Section 230 should be updated to account for recent technological changes.

### A. *Defective Products*

First are negligence and product liability claims involving defective websites, smartphone apps, or other internet technologies. As the internet has matured, software and other digital products have come to play an increasingly important role in society, often directly replacing physical products of yesteryear.[142] Unfortunately, not all digital products are well designed, and sometimes users suffer harms as a result. In the physical world, one available remedy is a negligence or product liability claim against the manufacturer.[143] But in the virtual world, those basic remedies are often barred by Section 230.[144]

---

139  Here and throughout the Article, online entities are typically described as if they perform only a single function that is or is not eligible for Section 230 immunity. In reality, an online entity can engage in a variety of activities for which it could face a multitude of legal claims. Eligibility for immunity is dependent on both the legal theory brought against it and the particular functionality being challenged. For example, although Facebook and eBay operate according to very different business models, both offer sales and messaging functionality, and their eligibility for immunity in a particular case would depend on the functionality being challenged and the theory alleged. *See* City of Chicago v. StubHub!, Inc., 624 F.3d 363, 366 (7th Cir. 2010) (reasoning that Section 230 would bar publication-related claims, but not the tax claim brought by the City of Chicago). *Compare* Chi. Laws.' Comm. for C.R. Under L., Inc. v. Craigslist, Inc., 519 F.3d 666, 668, 672 (7th Cir. 2008) (opinion of Easterbrook, C.J.) (finding the online classified ad site Craigslist immune from housing discrimination claim because its site hosted notices posted by others), *with* Fair Hous. Council of San Fernando Valley v. Roommates.com, LLC, 521 F.3d 1157, 1169–70 (9th Cir. 2008) (en banc) (opinion of Kozinski, C.J.) (rejecting Section 230 defense against housing discrimination claim because the website design included prepopulated dropdown box options).

140  47 U.S.C. § 230(c)(1); *see supra* Section I.C.

141  *See supra* Section I.C; *infra* Section II.B.

Consider recent lawsuits against Snapchat. Snapchat is a smartphone app that allows users to take temporary photos of themselves or others and to share them with friends.[145] Beyond ordinary photos, the Snapchat app includes various "filters" and "lenses" that users can apply to add premade graphic overlays to their photos.[146] One filter included with the Snapchat app was a speed filter, which was designed to calculate the user's current speed and superimpose that speed onto the user's photograph.[147] The Snapchat speed filter became the focus of two recent lawsuits by victims killed or seriously injured by teenage drivers traveling at speeds over 100 miles per hour while using Snapchat's speed filter.[148] In both cases, the plaintiffs al-leged that Snapchat was partly at fault for the accidents because the design of the app encouraged speeding and reckless driving by its users.[149]

---

142 *See* Ved Sen, *How Technology Is Reshaping Our Physical World*, IDG CONNECT (Apr. 8, 2014, 3:30 AM PDT), https://www.idgconnect.com/article/3580221/how-technology-is-reshaping-our-physical-world.html [https://perma.cc/W3E4-4R9U].

143 *See, e.g.*, Sean O'Kane, *Tesla Hit with Another Lawsuit over a Fatal Autopilot Crash*, VERGE (Aug. 1, 2019, 5:59 PM), https://www.theverge.com/2019/8/1/20750715/tesla-autopilot-crash-lawsuit-wrongful-death [https://perma.cc/52DD-ZXC6].

144 Courts have also not yet resolved whether software can constitute a "product" for purposes of product liability claims. *See* RESTATEMENT (THIRD) OF TORTS: PRODS. LIAB. § 19 cmt. d (AM. L. INST. 1998) (noting dispute and collecting "numerous commentators [who] have discussed the issue and urged that software should be treated as a product"); *see also* Stevens v. MTR Gaming Grp., Inc., 788 S.E.2d 59, 66–67 (W. Va. 2016) (declining to decide whether video lottery terminal software is a product because, regardless of whether or not it is a product, its use by a compulsive gambler did not make it defective); Blaisdell v. Dentrix Dental Sys., Inc., 284 P.3d 616, 619–21 (Utah 2012) (assuming without discussion that software could support a product liability claim); Winter v. G.P. Putnam's Sons, 938 F.2d 1033, 1036 (9th Cir. 1991) (suggesting in dicta that computer software might be a product for purposes of product liability law).

145 *See* Christine Elgersma, *Everything You Need to Know About Snapchat*, PHYS.ORG (June 18, 2018), https://phys.org/news/2018-06-snapchat.html [https://perma.cc/P95Y-BELB].

146 *See Snapchat Support: How to Use Filters*, SNAP INC., https://support.snapchat.com/en-US/article/geofilters [https://perma.cc/CKT4-TNB3]; *Snapchat Support: How to Use Lenses*, SNAP INC., https://support.snapchat.com/en-US/a/face-world-lenses [https://perma.cc/BUD8-VRVL].

147 Maynard v. Snapchat, Inc., 816 S.E.2d 77, 79 (Ga. Ct. App. 2018).

148 *See id.*; Lemmon v. Snap, Inc., 440 F. Supp. 3d 1103, 1105 (C.D. Cal. 2020), *appeal filed*, No. 20-55295 (9th Cir. Mar. 19, 2020); *see also* Katie Mettler, *Teen Took Snapchat Photos While Crashing Mercedes at 107 mph. Now Her Victim Has Sued Snapchat*, WASH. POST (Apr. 28, 2016, 7:13 AM MDT), https://www.washingtonpost.com/news/morning-mix/wp/2016/04/28/lawsuit-blames-snapchat-for-107-mph-crash-in-mercedes-caused-by-teen-girl-using-speed-filter [https://perma.cc/74H6-AV4D] (reporting on the *Maynard* case).

Were Snapchat a physical-world product, a negligence or product liability lawsuit would turn on familiar questions of tort law. Did Snapchat exercise reasonable care in avoiding foreseeable harm to the plaintiffs,[150] and was the product free from manufacturing, design, and labeling defects?[151] But because Snapchat operates in the virtual world, Section 230 provides it with heightened protection.[152]

In the virtual world, regardless whether an online entity sells a defective product or exercises reasonable care, it can assert immunity under Section 230 on the ground that it did not author the content at issue. For example, in *Lemmon v. Snap, Inc.*,[153] the plaintiff argued that Snapchat acted negligently by failing to redesign its speed filter despite numerous and widespread reports of motor vehicle accidents caused by teenage drivers using the speed filter while driving at high speeds to show off to their friends.[154] Rather than consider Snapchat's design and level of care, however, the court dismissed the case as barred by Section 230 because Snapchat was not the author of any content.[155] Courts have reached the same result in other cases involving alleged design defects in virtual products, reasoning that regardless of any negligence or defects, online entities are immune if they are not authors of the content in question.[156]

---

[149] *Maynard*, 816 S.E.2d at 81; *Lemmon*, 440 F. Supp. 3d at 1106.
[150] *See* RESTATEMENT (THIRD) OF TORTS: LIAB. FOR PHYSICAL & EMOTIONAL HARM §§ 3, 6, 19 (AM. L. INST. 2005).
[151] *See* RESTATEMENT (THIRD) OF TORTS: PRODS. LIAB. §§ 1–2 (AM. L. INST. 1998).
[152] *See* 47 U.S.C. § 230(c).
[153] 440 F. Supp. 3d 1103 (C.D. Cal. 2020).
[154] *See id.* at 1105–07.
[155] Specifically, the *Lemmon* court found that the Snapchat speed filter was not content at all, but rather a "content-neutral tool." *Id.* at 1109–11. Therefore, the court reasoned that Snapchat could not be the author of any content and was immune from liability under Section 230(c)(1), which immunizes online entities from claims if they are not an "information content provider" of the relevant content. *Id.* at 1109–13 (quoting 47 U.S.C. § 230(c)(1)). A Georgia state trial court held similarly in *Maynard v. McGee*, No. 16-SV-89, 2017 WL 384288, at *3 (Ga. State Ct. Jan. 20, 2017) (finding Snapchat immune under Section 230, reasoning that the Snapchat user, not Snapchat, created the content at issue). *But see* Maynard v. Snapchat, Inc., 816 S.E.2d 77, 79–82 (Ga. Ct. App. 2018) (reversing lower court and declining to extend immunity on ground that case did not depend on publication of third-party content); Maynard v. Snapchat, Inc., No. A20A1218, 2020 WL 6375424, at *3–4 (Ga. Ct. App. Oct. 30, 2020) (following failed Section 230 defense and remand, affirming dismissal for lack of duty of product manufacturer to protect against third-party misuses of product).
[156] *See, e.g.*, Herrick v. Grindr, LLC, 306 F. Supp. 3d 579, 590 (S.D.N.Y. 2018) (finding dating app immune under Section 230 despite inclusion of geolocation function and lack of

The point is not that Snapchat should or should not be liable for these motor vehicle accidents. Liability questions involving third-party actors are notoriously tricky. What these cases show, though, is how poorly suited Section 230 is for analyzing product-liability-type claims involving virtual products. First, content authorship is a poor test for potential culpability in this context. On the publication-centric internet of the past, nonauthors could safely be assumed to be benign content passthroughs; that is not true today.[157] That an entity does not author content should not immunize it from claims that its products are negligently designed.

Second, Section 230's poor fit with products liability claims cannot be explained as a necessary sacrifice to further the chief aims of the statute—preventing economically crippling duties to moderate and collateral censorship of user speech.[158] In a lawsuit involving an allegedly defective product design, a plaintiff typically alleges not that the defendant should have more thoroughly—and expensively—moderated user-created content, but that it should have incorporated some safety feature into its product.[159] Such a claim might threaten to impose extra costs on designers of virtual products, but not the insurmountable content-moderation burden Section 230 was designed to prevent. These ordinary tradeoffs between product safety, design, and development costs could be left to the same state negligence and product liability regimes that govern physical-world products.

### B. Online Marketplaces

Another category of problematic cases is those involving online marketplaces. Online marketplaces have replicated and, in many areas, supplanted traditional physical-world marketplaces as the preferred venues for advertising and for retail and commercial transactions.[160] That migration brought with it both good and bad: expanded product offerings; larger markets; and lightning-fast transactions; as well as defective products; negligence; race- and sex-based discrimination; and anticompetitive business practices. In the virtual

---

safety features which allowed user's ex-boyfriend to harass him), *aff'd*, 765 F. App'x 586 (2d Cir. 2019), *cert. denied*, 140 S. Ct. 221 (2019); Doe v. MySpace, Inc., 528 F.3d 413, 420–21 (5th Cir. 2008) (alleging that MySpace social networking site should have included age verification functionality and that such functionality would have prevented thirteen-year-old daughter's sexual assault by online predator).

157  *See supra* Section I.C.3.
158  *See* Wu, *supra* note 6, at 298–301, 309–13.
159  *See, e.g.*, *MySpace*, 528 F.3d at 420–21.
160  *See supra* Section I.C.3.

world, though, Section 230 often bars victims of unsavory marketplace elements from seeking relief.

For example, in 2016, the U.S. Court of Appeals for the First Circuit considered *Doe v. Backpage.com, LLC*,[161] which involved an action brought against Backpage by three underage girls who had become victims of sex trafficking and sold on the website.[162] According to the complaint, when Backpage's competitor, Craigslist, closed its "Adult Services" section in 2010 due to sex trafficking concerns, Backpage intentionally enhanced the "Escorts" section of its website to maximize its profits by making sex trafficking easier.[163] The plaintiffs alleged that Backpage had "deliberate[ly] structure[ed] . . . its website to facilitate sex trafficking" and that it had "tailored its posting requirements" and established "rules and processes governing the content of advertisements" in a way that encouraged and facilitated sex trafficking.[164] For example, Backpage removed postings made as part of law enforcement sting operations and removed metadata from escort photographs to limit their usefulness to law enforcement agencies.[165] The plaintiffs alleged that Backpage made such moves to profit from sex traffickers' use of the website.[166] The "Adult" section was the only section of Backpage's site that charged a posting fee.[167] And for an extra fee, Backpage allowed users to post "Sponsored Ads" that appeared on the right hand side of every page in the "Escorts" section and included a picture of the advertised individual as well as her location and availability.[168]

The underage girls who had become victims of sex trafficking publicized on Backpage asserted what was essentially a civil conspiracy claim[169] against Backpage under the Trafficking Victims Protection Reauthorization Act ("TVPRA") of 2017,[170] which includes a private right of action against anyone who "knowingly benefits . . . from participation in a venture which that person knew or should have known has engaged in an act [of sex trafficking]."[171] The plaintiffs al-leged that Backpage intentionally structured its website to aid the sex trafficking operations that had victimized them and to thereby maximize Backpage's profits—the vast majority of which came from the "Escorts" section of its webpage.[172]

---

[161] 817 F.3d 12 (1st Cir. 2016).

[162] *Id.* at 16.

[163] *See* Second Amended Complaint at ¶¶ 29–51, Doe v. Backpage.com, LLC, 104 F. Supp. 3d 149 (D. Mass. 2015) (No.14-13870) [hereinafter Backpage Complaint].

[164] *Backpage.com*, 817 F.3d at 16–17; *see* Backpage Complaint, *supra* note 163, at ¶¶ 52–59.

[165] *Backpage.com*, 817 F.3d at 16; Backpage Complaint, *supra* note 163, at ¶¶ 40, 51.

[166] Backpage Complaint, *supra* note 163, at ¶ 45.

[167] *Backpage.com*, 817 F.3d at 17; Backpage Complaint, *supra* note 163, at ¶ 43.

[168] *Backpage.com*, 817 F.3d at 17; Backpage Complaint, *supra* note 163, at ¶¶ 53–59.

[169] *See* Backpage Complaint, *supra* note 163, at ¶¶ 108–14.

[170] 22 U.S.C. §§ 7101–7114.

[171] 18 U.S.C. § 1595(a).

In the physical world, the court might have wrangled with difficult questions about Backpage's alleged level of assistance to the sex traffickers and whether it knowingly participated in or benefited from a joint venture. But under Section 230's alternate, virtual-world regime, the case was resolved by a straightforward application of the statute's text.[173] Backpage was immune from civil[174] liability because all claims related to the sex trafficker's escort listings on Backpage involved "information provided by [someone else]:"[175] "Whatever Backpage's motivations, those motivations do not alter the fact that the complaint premises liability on the decisions that Backpage is making as a publisher with respect to third-party content."[176] "[S]hield[ing] Backpage from liability here is congruent with the case law elsewhere."[177]

The law treats online and offline marketplaces differently. Under Section 230, traditional common law or statutory questions of intent, negligence, foreseeability, causation, and the like, are often irrelevant to whether an online marketplace will be liable for the harms suffered by its users. Applying Section 230, courts have thus dismissed claims against online marketplaces for claims as varied as negligence,[178] un-fair competition,[179] retailer product liability,[180] and housing discrimination in violation of the Fair Housing Act.[181] Even where an online marketplace's contribution to the wrongdoing is significant—in Backpage's case, intentionally designing its website to help sex traffickers evade detection[182]—and even where a physical-world entity would be liable for engaging in the same conduct, an online marketplace will be immune from liability as long as its role in the scheme does not involve the creation of content.[183]

---

[172] *Backpage.com*, 817 F.3d at 16.

[173] *Id.* at 22.

[174] Importantly, although it bars victims' civil claims for recovery, Section 230 does not prevent criminal prosecution. *See* 47 U.S.C. § 230(e); Tom Jackman, *Backpage CEO Pleads Guilty in Three States*, WASH. POST, Apr. 14, 2018, at A3 (reporting criminal prosecutions of Backpage officials for conspiracy to facilitate prosecution). Whether criminal prosecution of Backpage and Congress's later amendment of Section 230 to permit civil claims against online entities benefitted sex workers is an open question. Criminalization and other legal barriers to sex work may harm sex workers by forcing them to work in secret, more dangerous environments. *See* Anna North, *Sex Workers Are in Danger. Warren and Sanders Are Backing a Bill That Could Help*, VOX (Dec. 17, 2019, 12:20 PM), https://vox.com/identities/2019/12/17/21024859/sex-work-bernie-sanders-elizabeth-warren-fosta [https://perma.cc/6L56-4CSS].

[175] 47 U.S.C. § 230(c).

[176] *Backpage.com*, 817 F.3d at 21.

[177] *Id.* In direct response to the Backpage decision, Congress enacted the Allow States and Victims to Fight Online Sex Trafficking Act of 2017 ("FOSTA"), Pub. L. 115-164, 132 Stat. 1253 (2018) (codified as amended in scattered sections of 18 & 47 U.S.C.), which amended Section 230 to permit certain sex trafficking claims against online entities. *See infra* Section III.A.1.

[178] *See, e.g.*, Dyroff v. Ultimate Software Grp., Inc., 934 F.3d 1093, 1097–101 (9th Cir. 2019) (affirming dismissal of negligence and other claims against social networking platform Experience Project for claims related to a man's death following a drug deal among the platform's users, where decedent had participated in an anonymous heroin-related forum, the site was de-

---

signed to allow anonymous users, and the platform's algorithm connected the drug dealer and the decedent by creating topical discussion groups).

179    *See, e.g.*, Marshall's Locksmith Serv., Inc. v. Google, LLC, 925 F.3d 1263, 1265–66, 1272 (D.C. Cir. 2019) (affirming dismissal of locksmith companies' Lanham Act, Pub. L. No. 79-489, 60 Stat. 427 (1946) (codified as amended in scattered sections of 15 U.S.C.), false advertising claims, which alleged that Google and other online advertising platforms had "conspired to 'flood the market' of online search results with information about so-called 'scam' locksmiths, in order to extract additional advertising revenue" from truly local locksmiths); *see also* Caraccioli v. Facebook, Inc., 700 F. App'x 588, 590 (9th Cir. 2017) (affirming dismissal of unfair competition claim alleging Facebook violated its own terms of service to plaintiff's detriment by failing to block obscene videos of plaintiff posted to the service by an unknown person).

180    *See, e.g.*, Oberdorf v. Amazon.com, Inc., 930 F.3d 136, 151–53 (3d Cir. 2019) (dismissing product liability claim against Amazon for defective product warnings to the extent it related to information provided on Amazon product listing page), *reh'g en banc granted and vacated*, 936 F.3d 182 (3d Cir. 2019), *certifying different question to Pa. Sup. Ct.*, 818 F. App'x 138 (3d Cir. 2020); Gartner v. Amazon.com, Inc., 433 F. Supp. 3d 1034, 1045 (S.D. Tex. 2020) (same). *But see* Bolger v. Amazon.com, LLC, 267 Cal. Rptr. 3d 601, 605, 626 (Cal. Ct. App. 2020) (allowing product liability claim to proceed against Amazon because "it was pivotal in bringing the product . . . to the consumer" and rejecting Section 230 defense on ground that action "is based on Amazon's own conduct" and "not the content of [the third-party seller's] product listing").

181    42 U.S.C. §§ 3601–3619; *see* Chi. Laws.' Comm. for C.R. Under L., Inc. v. Craigslist, Inc., 519 F.3d 666, 668, 672 (7th Cir. 2008) (finding Craigslist immunized from claims brought under the Fair Housing Act, which forbids a person "[t]o make, print, or publish, or cause to be made, printed, or published any . . . statement" that indicates discriminatory preference regarding the sale or rental of a dwelling (quoting 42 U.S.C. § 3604(a) (alteration in original))). *But see* Fair Hous. Council of San Fernando Valley v. Roommates.com, LLC, 521 F.3d 1157, 1169–70 (9th Cir. 2008) (en banc) (rejecting Section 230 defense against Fair Housing Act claim on ground that design of website to include dropdown box options made Roommates.com a content creator).

182    Doe v. Backpage.com, 817 F.3d 12, 16 (1st Cir. 2016).

183    The Ninth Circuit's influential decision in *Roommates.com* initially appeared poised to narrow immunity by defining content creators to include websites that assist their users in developing content. *See Roommates.com*, 521 F.3d at 1166. Subsequent courts have continued to apply immunity broadly, however, requiring plaintiffs to show an entity's material contribution by "specifically encourag[ing] development of what is offensive about the content" before an entity can be found to have created or developed content. FTC v. Accusearch Inc., 570 F.3d 1187, 1199 (10th Cir. 2009); *see* Malwarebytes, Inc. v. Enigma Software Grp. USA, LLC, No. 19-1284, 2020 WL 6037214 (U.S. Oct. 13, 2020) (Thomas, J., statement respecting the denial of certiorari) (noting that courts have narrowly construed content creation "to cover only substantial or material edits and additions"); *see also* Sylvain *supra* note 9, at 258 (explaining that "[i]n practice" the

## C. Volitional Wrongs

A final category of problematic cases is those involving volitional wrongs. Volitional is not meant to imply a specific mental state. Instead, it refers more generally to any conduct that is wrongful because of some knowledge or intent on the part of the defendant beyond that of a passive intermediary. Because Section 230 was designed for a world of passive content intermediaries, it does not consider an online entity's knowledge or intent.[184] Passive intermediaries merely convey the content of others, and holding them responsible for that content could impose a crippling content-moderation burden and spur them to censor user content.[185] Thus, under Section 230, if the entity is not an author of content, it is immune from liability for harms that result from that content.[186] But online entities often do more than act as passive intermediaries. And because Section 230's trigger for withholding immunity is content authorship, online entities can be immune under the law despite knowledge of or even intent to cause harms.

For example, in *Barnes v. Yahoo!, Inc.*,[187] the Ninth Circuit considered a claim brought by a plaintiff whose former boyfriend sought revenge for their breakup by creating fake profiles under her name in which he included her contact information and nude photographs that had been taken without her knowledge.[188] Soon after, the plaintiff received a barrage of emails, phone calls, and in-person visits from men she did not know, all with the expectation of sex.[189] The plaintiff contacted Yahoo to remove the profiles, and Yahoo directed her to submit an official removal request by mail with a copy of her photo ID.[190] After more than a month and sending two more mailings, the profiles remained online, and the plaintiff had received no response.[191] Finally, one day before a local news station broadcasted a report regarding the

---

material contribution standard "makes legal challenges to intermediaries' designs especially difficult to win").

[184] *See* 47 U.S.C. § 230(c)(1); *see also, e.g.*, Daniel v. Armslist, LLC, 926 N.W.2d 710, 726 (Wis. 2019) (explaining that plaintiff's allegation that Armslist knew its website was used for illegal gun sales "does not change the result" because Section 230 "contains no good faith requirement" and "courts do not allow allegations of intent or knowledge to defeat a motion to dismiss"), *cert. denied*, 140 S. Ct. 562 (2019).

[185] *See* Wu, *supra* note 6, at 295–96.
[186] 47 U.S.C. § 230(c)(1).
[187] 570 F.3d 1096 (9th Cir. 2009).
[188] *See id.* at 1098.
[189] *Id.*
[190] *Id.*
[191] *Id.*

incident, the plaintiff received a call from Yahoo's Director of Communications, who assured her that Yahoo would remove the profiles.[192] Two more months passed with no change, at which point the plaintiff filed a lawsuit against Yahoo alleging its failure to remove the profiles after promising to do so constituted a "negligent undertaking."[193] The profiles were taken down shortly after the plaintiff's suit was filed.[194]

Had the case been based on Yahoo's publication of the photos—for example, a defamation or invasion of privacy claim—it would be a prime example of the sort of claim Section 230 is designed to protect against: Yahoo could not reasonably review each profile posted to its system, and a legal regime that required it to do so could lead it to shut down or curtail user speech. But in this case, Yahoo did more than fail to detect an unlawful third-party post. It expressly acknowledged the unlawful profiles, promised that it would remove them, and then carelessly failed to follow through.[195] Given its knowledge, no mass filtering or needle-in-a-haystack search was required. Nonetheless, Section 230 barred the plaintiff's claim because Yahoo's actions related to content authored by a third party.[196]

In the physical world, tort law distinguishes between two types of cases. On one hand, bookstores, libraries, and other entities that merely provide access to content authored by others are not ordinarily subject to defamation liability, for "it would be an unreasonable burden to require [such entities] to make the investigation necessary"[197] to evaluate all of the content they distribute. On the other hand, however, such entities can be subject to liability if they know or have reason to know that particular content is defamatory.[198] By treating these two situations differently, defamation law provides liability protection to entities that need it—those that would otherwise face an unreasonable content moderation burden—while still allowing claims against entities whose specific knowledge removes the burden to moderate content.[199]

---

192  *Id.* at 1098–99.
193  *Id.* at 1099.
194  *Id.*
195  *See id.*
196  *Id.* at 1102–03.
197  Restatement (Second) of Torts § 581 cmt. f (Am. L. Inst. 1965).
198  *Id.* § 581.
199  *Id.* § 581 cmts. e–f (explaining that a bookstore or library "is not required to examine [its publications] to discover whether they contain anything of a defamatory character" but can face liability if it "knows or has reason to know that the [content] is libelous"); *see also* Skorup &

But Section 230 includes no distinction based on scienter and does not limit immunity to those instances where a claim would impose a content-moderation burden on the defendant. Instead, Section 230 equally protects against all claims as long as an online entity is not the author of the content in question.[200] Under this principle, online entities are absolutely immune from liability even if they are aware of or even intend to cause the harms that result from the content they distribute on their platforms.[201]

### III. Rebooting Internet Immunity

As detailed in Part II, Section 230's content-authorship immunity rule is poorly tailored to the modern internet, which includes not only content distributors but also many other entities whose primary function is to provide real-world goods and services. Despite the mismatch, however, strong arguments remain today for Section 230's protections with respect to important portions of the internet. Although content-monitoring capacity has improved drastically with the help of artificial intelligence, the Twitters and Facebooks of the world still experience great difficulty in reliably reviewing the mass of content that flows through their systems,[202] and there are good reasons why it might be undesirable for them to do so. Holding online entities liable for user-authored content would encourage them to censor user speech on their platforms and stifle free expression on the internet.[203] Accordingly, when Section 230 immunizes such entities from claims

---

Huddleston, *supra* note 96, at 638–46 (detailing defamation and, in particular, mass media law's evolution throughout the 20th century away from strict liability and toward fault-based rules).

200 *See* 47 U.S.C. § 230(c)(1) ("No provider or user of an interactive computer service shall be treated as the publisher or speaker of any information provided by another information content provider.").

201 *See, e.g.*, Doe v. Backpage.com, 817 F.3d 12, 21–24 (1st Cir. 2016) (dismissing sex trafficking conspiracy claim against website despite allegations it deliberately implemented measures to make sex trafficking on its platform easier because "[w]hatever Backpage's motivations, those motivations do not alter the fact that the complaint premises liability on . . . third-party content"); *Barnes*, 570 F.3d at 1098–1103; Doe v. Am. Online, Inc., 783 So. 2d 1010, 1018 (Fla. 2001) (finding Section 230 to bar action against AOL for violating Florida statutes prohibiting distribution of child pornography despite allegation that AOL was aware that a particular user of its service was transmitting unlawful photographs and yet declined to intervene); Batzel v. Smith, 333 F.3d 1018, 1035 (9th Cir. 2003) (noting operator of museum security website who received an email from an individual who suspected plaintiff was a Nazi art thief, which email operator then independently edited and posted online without verification, may be immune under Section 230(c)(1) because he was not the author of the original email).

202 *See* John Naughton, *Facebook's Burnt-Out Moderators Are Proof that It Is Broken*, Guardian (Jan. 6, 2019, 2:00 AM), https://www.theguardian.com/commentisfree/2019/jan/06/proof-that-facebook-broken-obvious-from-modus-operandi [https://perma.cc/2Z5R-XK66].

203 *See supra* Section I.C.2.

related to user content, it continues to serve a purpose that is reasonably close to what the Congress that enacted it had in mind.[204]

Given the critical role it plays, before discussing possible reforms, an important preliminary question is whether Section 230 can be reformed without jeopardizing the important protections it provides. This Part begins by addressing the risks of changes to Section 230's protections, before moving on to discuss how reforms might be structured to avoid undermining the law's key protections.

### A. Updating Section 230 Safely

For decades now, Section 230 has been treated as sacred. Industry leaders, joined by prominent legal scholars, have presented a united voice, warning that changes to the statute would undermine the American tech industry and the internet as we know it.[205] But as the internet has continued to evolve beyond publication into an ever more complete virtual world, Section 230's mismatch with the modern internet has become hard to ignore. Indeed, views have shifted to the point where Section 230 now finds itself in the crosshairs of legal scholars and both major political parties. Reform proposals can be grouped into three categories.

#### 1. Recent Proposals for Reform

First, one group of proposals would remove immunity protection where an entity intentionally facilitates unlawful conduct. Some such proposals focus on stripping immunity from the most egregious offenders: those online entities that intentionally facilitate criminal behavior. For example, a recent Department of Justice review of Section 230 concluded that a "Bad Samaritan" carve out[206] should be added to the statute to ensure that online entities that purposefully solicit "third parties to sell illegal drugs to minors, exchange child sexual abuse material," or engage in other unlawful activities through their platforms "do not benefit from Section 230's sweeping immunity at the expense of their victims."[207] In a similar vein, Congress specifically targeted intentional unlawful conduct in 2018 when it enacted the Allow States and Victims to Fight Online Sex Trafficking Act of 2017 ("FOSTA")[208] in response to the First Circuit's *Backpage.com* decision.[209] Among other things, FOSTA authorized private actions against websites for facilitating sex trafficking by "publishing information designed to facilitate sex trafficking"[210] and prevented websites from invoking Section 230(c)(1) to escape liability by adding subsection (e)(5), which specifically excludes violators from protection.[211]

---

[204] *See* 47 U.S.C. § 230(a)–(b).

[205] *See, e.g.*, Elliot Harmon, *FOSTA Would Be a Disaster for Online Communities*, Elec. Frontier Found.: Deeplinks (Feb. 22, 2018), https://www.eff.org/deeplinks/2018/02/fosta-would-be-disaster-online-communities [https://perma.cc/9FAF-EQVQ]; Derek E. Bambauer, *How Section 230 Reform Endangers Internet Free Speech*, Brookings: Tech Stream (July 1, 2020), https://www.brookings.edu/techstream/how-section-230-reform-endangers-internet-free-speech/ [https://perma.cc/EL6X-ZSAR]; Jeff Kosseff, *Defending Section 230: The Value of Intermediary Immunity*, 15 J. Tech. L. & Pol'y 123, 126 (2010).

[206] *See* DOJ Section 230 Recommendations, *supra* note 22, at 14. In recommending a Bad Samaritan carve out, DOJ referenced *Daniel*, noting that, in that case, the website that facilitated the sale of a firearm to a prohibited person "was immune under Section 230, despite allegations that [the] website was intentionally designed with the specific purpose of skirting federal firearm laws." *Id.*

Other reform proposals also look to an entity's intentionality, but would exclude from immunity not only those entities that intentionally facilitate unlawful conduct, but also those that intentionally form cooperative—and typically profitable—relationships with third-party wrongdoers.[212] For example, in her forthcoming article, Agnieszka McPeak argues that Section 230 should incorporate joint enterprise liability theory to ensure that sharing-economy facilitators like Uber and Amazon cannot avoid liability for harms caused by their businesses by characterizing themselves as mere online intermediaries between customers and freelance workers or third-party sellers.[213] Such approaches expand potential liability beyond the most egregious abuses, but still limit the moderation burden they impose on entities by permitting liability only where an entity has a heightened mental state of intentionality, either as to the wrongful conduct itself, or the cooperative relationship out of which it arose.[214]

---

[207] DOJ Section 230 Recommendations, *supra* note 22, at 14; *see also* Tushnet, *supra* note 6, at 1009–10 (arguing that Section 230 immunity should include limits on an intermediary's power to control speech, including where an online intermediary deliberately induces the creation of unlawful content for financial gain); Stacey Dogan, *Principled Standards vs. Boundless Discretion: A Tale of Two Approaches to Intermediary Trademark Liability Online*, 37 COLUM. J.L. & ARTS 503, 507–08 (2014) (in trademark infringement context, discussing benefits of liability rule focused on intermediaries' intentional solicitation of and profit from infringing content).

[208] Pub. L. No. 115-164, 132 Stat. 1253 (2018) (codified as amended in scattered sections of 18 & 47 U.S.C.).

[209] *See supra* Section II.B; Tom Jackman, *Trump Signs 'FOSTA' Bill Targeting Online Sex Trafficking, Enables States and Victims to Pursue Websites*, WASH. POST (Apr. 11, 2018, 11:41 AM MDT), https://www.washingtonpost.com/news/true-crime/wp/2018/04/11/trump-signs-fosta-bill-targeting-online-sex-trafficking-enables-states-and-victims-to-pursue-websites [https://perma.cc/Z5E6-QJSN].

[210] 132 Stat. at 1255.

[211] *See* 47 U.S.C. § 230(e)(5).

[212] *See* McPeak, *supra* note 23; Sylvain, *supra* note 9, at 276–77 (suggesting online immunity doctrine should consider entities' intentional solicitation and sale or use of user data); Dickinson, *supra* note 85, at 877–81 (arguing that Section 230 should be read to incorporate tort law theories of vicarious liability including concert of action).

[213] *See* McPeak, *supra* note 23.

Second, another set of proposals would deny Section 230 immunity where an online entity has actual knowledge of unlawful content on its platform. Reform proposals of this sort are modeled after the Digital Millennium Copyright Act's ("DMCA")[215] notice-and-takedown procedure[216] and would deny Section 230 immunity to entities that continue to provide access to unlawful content or facilitate unlawful behavior despite actual knowledge of the content or behavior.[217] Such a rule would be analogous to the common-law principle, discussed earlier,[218] under which bookstores and other media distributors lose immunity from defamation claims if they have actual or constructive knowledge of defamatory content. And, as with physical-world distributors, notice- or knowledge-based limitations on immunity avoid imposing a burden on entities to proactively screen all third-party content.

Third, a final group of proposals would focus on an entity's platform-policing practices in general, rather than its response to the particular piece of content that precipitated a plaintiff's alleged harms. Danielle Citron and Benjamin Wittes, for example, have proposed amending Section 230 to require that an entity take "reasonable steps to prevent or address unlawful uses of its services" to be eligible for immunity.[219] Making immunity contingent on entities' practices as a whole could encourage industry organizations to develop best practices for policing online content and foster a business culture more attuned to the harms caused by the technologies they deploy. Reasonable policing practices could be defined either abstractly, via a reasonableness standard like that proposed by Citron and Wittes, or could be particularized via statutory or regulatory mandate.[220] A reasonableness standard would provide courts flexibility to adapt Section 230 for new contexts and different types of entities,[221] whereas particularized regulations would provide entities greater certainty ex ante regarding whether their practices would qualify them for immunity.

---

[214] *Cf.* DOJ Section 230 Recommendations, *supra* note 22, at 14–15 (recommending "Bad Samaritan" carve out that includes a "heightened *mens rea*, such as 'purposefully,' under which platforms that accidentally or even negligently facilitate unlawful behavior would not lose immunity" so that the carve out would not "impose a burden on platforms to proactively screen all third-party content").

[215] Digital Millennium Copyright Act, Pub. L. 105-304, 112 Stat. 2860 (codified as amended in scattered sections of 17 & 28 U.S.C.).

[216] *See* 17 U.S.C. § 512(c)(1) (defining procedure for copyright owners to notify online service providers of copyright-infringing material and granting providers safe harbor from infringement liability if they remove the allegedly infringing material from their systems).

[217] *See* Citron & Wittes, *supra* note 9, at 455–56 (arguing immunity should be limited to online entities that, when warned, take reasonable steps to protect against illegal activity); Edelman & Stemler, *supra* note 9, at 193 (proposing immunity should be denied, among other times, when entities were on actual notice of a specific pattern or problem); Michael L. Rustad & Thomas H. Koenig, *Rebooting Cybertort Law*, 80 WASH. L. REV. 335 (2005) (suggesting no immunity where an entity has actual notice of ongoing unlawful activity); Tushnet, *supra* note 6, at 1010 (noting immunity could be denied where an entity refuses to remove content if the original speaker has conceded liability); *see also* DOJ Section 230 Recommendations, *supra* note 22, at 17–18 (proposing carve out from Section 230 immunity where an entity has actual knowledge of criminal material and citing *M.A. ex rel. P.K. v. Vill. Voice Media Holdings, LLC*, 809 F. Supp. 2d 1041, 1050 (E.D. Mo. 2011), which held Backpage immune under Section 230 despite allegation that entity had actual knowledge of child sex trafficking on its website).

[218] *See supra* Section I.C.2.

### 2. *Reform Risks and Limitations*

Although momentum for change is building, legislators have been rightly cautious. Because the statute is so broad, changes could have major effects on the tech industry. Some fear that altering Section 230's protections might even endanger the United States' position as a worldwide innovation leader.[222] Successfully reforming Section 230 requires navigating at least three potential obstacles: the difficulty of creating a single, unitary framework to govern all internet conduct; the risk of increased uncertainty and compliance costs associated with a more flexible rule; and the risk of increased litigation expenses in the technology industry.

The first obstacle is the difficulty of crafting a single rule or set of rules to govern the entire virtual world. On the publication-centric internet of 1996, internet wrongdoers could be identified with a simple test: Did the entity author content?[223] Designing an immunity rule for today's internet is much more difficult, where it must govern not only traditional content intermediaries but also the full gamut of other human social and economic activity, including digital product manufacturers, online marketplaces, and the harms that come with them. Given the broad range of wrongful conduct on the modern internet, reform proposals tend to focus on a specific category of wrongdoing— for example, websites that intentionally facilitate criminal activity[224]— or rely on reasonableness or multifactor tests that are flexible enough to apply in a variety of contexts.[225] Even such targeted and flexible approaches, though, have their limitations. For example, creating a carve out from Section 230 immunity for intentional or knowing wrongdoing would still leave unaddressed Section 230's special treatment of online entities facing reasonable care or strict liability claims, such as ordinary negligence or products liability. It is difficult for a single rule to reproduce the full range of standards applicable in all areas of the law.

---

[219] *See* Citron & Wittes, *supra* note 25, at 419.

[220] Two Brobdingnagianly titled bills proposed this year would take the particularized regulatory approach. First, the EARN IT Act of 2020 would amend Section 230 to allow civil suits against companies that recklessly distribute child pornography while simultaneously providing for a set of best practices to be developed by a National Commission on Online Child Sexual Exploitation Prevention to be established for the purpose. *See* Eliminating Abusive and Rampant Neglect of Interactive Technologies ("EARN IT") Act of 2020, S. 3398, 116th Cong. (2020). Second, the SHOP SAFE Act of 2020 would introduce a similar best practices safe harbor from liability for trademark infringement. *See* Stopping Harmful Offers on Platforms by Screening Against Fakes in E-commerce ("SHOP SAFE") Act of 2020, H.R. 6058, 116th Cong. (2020).

[221] For example, internet service providers and social networks that provide access to millions of user posts per day cannot plausibly respond to all complaints of abuse immediately. But they might be able to deploy algorithms that can automatically detect and remove certain types of unlawful content, such as child pornography or copyright infringing content that has previously been deemed unlawful. *See* Klonick, *supra* note 5, at 1635–37. The duty of care would vary by entity type and size and require more of online entities as technology improves.

[222] *See* Goldman, *supra* note 6, at 33–34; Kosseff, *supra* note 12, at 278–80. *But see* Paul Ohm, *We Couldn't Kill the Internet if We Tried*, 130 Harv. L. Rev. F. 79 (2016).

A second obstacle to reform is closely related to the first. Internet immunity doctrine can be made more flexible by introducing carve outs or by adding a reasonableness or mens rea requirement, but adding that flexibility also introduces greater uncertainty ex ante and increases compliance costs.[226] One way to see the problem is as an instance of the familiar question of rules versus standards: Should internet immunity be governed by a highly specific, bright line rule or by a more open-ended standard?[227] The choice is subject to the usual tradeoffs. Rules promote uniformity, predictability, and low decision costs at the expense of rigidity, whereas standards permit nuance, flexibility, and case-specific deliberation at the expense of uncertainty and high decision costs.[228] In the internet immunity context, uncertainty is especially dangerous because it undermines Section 230's core aims—relieving online entities from the immense content-moderation burden that could spur them to shutter their operations or censor user speech to avoid legal liability.[229] A major risk of Section 230 reform is that adding the flexibility required to govern the full range of virtual-world activity will undermine the very goals that the statute was designed to achieve.[230]

---

[223] *See supra* Sections I.C.3, II.A–.C.

[224] *See, e.g.*, DOJ Section 230 Recommendations, *supra* note 22, at 14–18 (proposing several specific carve outs from Section 230 immunity to exclude intentional facilitation of criminal wrongdoing; actual knowledge of content that violates federal criminal law or court judgments; and failure to take precautions to prevent terrorism, child sex abuse, or cyberstalking).

[225] *See, e.g.*, Citron & Wittes, *supra* note 25, at 419 (proposing that Section 230 be amended to require entities take "reasonable steps" to prevent or address unlawful uses of their services); McPeak, *supra* note 23 (manuscript at 34) (suggesting that courts incorporate tort law's joint enterprise theory of vicarious liability to add nuance to the "current binary classification" of entities as either intermediaries or content providers).

[226] Eric Goldman makes this point in a recent essay comparing Section 230 and First Amendment doctrine. *See* Goldman, *supra* note 6, at 45–46.

[227] *See* Louis Kaplow, *Rules Versus Standards: An Economic Analysis*, 42 Duke L.J. 557 (1992); Duncan Kennedy, *Form and Substance in Private Law Adjudication*, 89 Harv. L. Rev. 1685, 1687–1713 (1976); Antonin Scalia, *The Rule of Law as a Law of Rules*, 56 U. Chi. L. Rev. 1175 (1989); Pierre Schlag, *Rules and Standards*, 33 UCLA L. Rev. 379 (1985); Kathleen M. Sullivan, *Foreword: The Justices of Rules and Standards*, 106 Harv. L. Rev. 22 (1992); Cass R. Sunstein, *Problems with Rules*, 83 Calif. L. Rev. 953 (1995).

A potential approach to mitigate any reform's impact on Section 230's content-moderation and collateral censorship protections is to limit the reform to claims that require a heightened mental state such as "purposely" or "knowingly" rather than simple negligence. In theory, a heightened mental state requirement would relieve entities from any obligation to proactively screen third-party content because they could face liability only for content they intended or of which they were at least aware.[231] The reality, however, would be more complicated. First, courts deem corporate defendants to possess "knowledge" of information even if it is buried within their files or email systems and unknown to the entities' actual decision makers.[232] The burden of assessing what they collectively know or intend could lead organizations to adopt more stringent censorship policies to minimize the risk of liability from information technically within their spheres of knowledge. Second, tying liability to knowledge could have undesirable effects on entities' use of content-flagging mechanisms. Knowledge-based liability might chill user speech by leading platforms to automatically take down, without investigation, any content even reported as objectionable—a problem known as the "heckler's veto."[233]

---

[228] *See* Kaplow, *supra* note 227, at 596–620.

[229] *See supra* Section I.C.2.

[230] *See* Goldman, *supra* note 6, at 45–46 ("Section 230's agnosticism about defendant scienter is a key element of its success.").

[231] DOJ Section 230 Recommendations, *supra* note 22, at 14.

[232] *See* RESTATEMENT (THIRD) OF AGENCY § 5.03 (AM. L. INST. 2006); 3 WILLIAM MEADE FLETCHER, FLETCHER CYCLOPEDIA OF THE LAW OF CORPORATIONS § 790 (2020) ("[T]he general rule is . . . that a corporation is charged with constructive knowledge, regardless of its actual knowledge, of all material facts of which its officer or agent . . . acquires knowledge . . . even though the officer or agent does not in fact communicate the knowledge to the corporation.").

[233] *See generally* Brett G. Johnson, *The Heckler's Veto: Using First Amendment Theory and Jurisprudence To Understand Current Audience Reactions Against Controversial Speech*, 21 COMMC'NS L. & POL'Y 175, 175–81 (2016) (discussing the concept of the heckler's veto, whereby an individual is able to restrict another's freedom to speak by filing complaints against, shouting

Alternatively, tying liability to knowledge could even lead some entities to remove content-flagging mechanisms altogether so as to never become aware of objectionable content in the first place.

Various designs might overcome these difficulties. For example, replacing Section 230's content-authorship test with a different but similarly bright line rule would limit uncertainty-driven censorship of user speech, as would employing an objective standard of care[234] rather than a mental state inquiry. Regardless of the particular approach, however, the key point is that because they can alter existing incentives for entities to police online content, reforms that add flexibility to Section 230's immunity rule must be carefully designed to ensure they do not undermine the statute's speech-enhancing objectives.

Finally, a third obstacle to reform is the effect that changes to internet immunity doctrine could have on litigation expenses in the tech industry. As currently designed, Section 230 provides online entities with both substantive and procedural protections beyond those of physical-world entities.[235] Substantively, online entities are protected from any legal obligation to review and moderate third-party con-

---

down, heckling, threatening, or otherwise harassing the speaker); *see also* Reno v. ACLU, 521 U.S. 844, 880 (1997) (invalidating portions of the CDA, because, among other reasons, the requirement not to communicate indecent speech to "'specific person[s]' . . . would confer broad powers of censorship, in the form of a 'heckler's veto,' upon any opponent of indecent speech" (quoting 47 U.S.C. § 223(d))); Rory Lancman, *Protecting Speech from Private Abridgement: Introducing the Tort of Suppression*, 25 Sw. U. L. Rev. 223, 253–55 (1996) (discussing the origin of the "heckler's veto" concept).

234   The proposed EARN IT Act of 2020, for example, would establish a national commission tasked with crafting best practices for online entities to prevent the distribution of child pornography. *See* Eliminating Abusive and Rampant Neglect of Interactive Technologies ("EARN IT") Act of 2020, S. 3398, 116th Cong. § 3 (2020). Compliance with those best practices would entitle entities to safe harbor from civil suits. *See id.* By tying safe harbor to a clear set of best practices, the EARN IT Act would reduce the risk of collateral censorship. For further discussion of the merits and risks of the proposal, compare Riana Pfefferkorn, *The EARN IT Act Threatens Our Online Freedoms. New Amendments Don't Fix It.*, Stan.: Ctr. for Internet & Soc'y (July 6, 2020, 11:45 AM), https://cyberlaw.stanford.edu/blog/2020/07/earn-it-act-threatens-our-online-freedoms-new-amendments-don%E2%80%99t-fix-it [https://perma.cc/YHG4-MGF8] (arguing that the Act would lead to overcensorship of user speech and that providers might be hesitant to employ end-to-end encryption technologies despite their express protection in the most recent version of the bill, because of the potential for protected litigation), with Stewart Baker, *A New Twist in the Endless Debate over End-to-End Encryption*, Volokh Conspiracy (Feb. 11, 2020, 3:11 PM), https://reason.com/2020/02/11/a-new-twist-in-the-endless-debate-over-end-o-end-encryption [https://perma.cc/5J5Z-XJVP] (acknowledging concerns regarding end-to-end encryption, but noting that the legal system ordinarily requires entities to internalize the costs of the harms their products cause to the public).

235   *See* Goldman, *supra* note 6, at 36, 39.

tent.236 Procedurally, those substantive protections are enhanced by Section 230's simplicity, which does not consider mental state and requires only that the defendant did not author the content in question.237 The showing required to make out a Section 230 defense is so simple that courts are often able to resolve claims before trial, either at the pleading stage or by summary judgment,238 which reduces online entities' litigation related expenses.

Reforms should thus be undertaken cautiously and designed to avoid unnecessarily increasing the cost of litigation. However, unlike changes to Section 230's substance—that is, changes that would require entities to moderate user-created content—measured changes to Section 230's procedural efficiency would increase costs but not threaten the viability of internet entities' basic business models. Comparable physical-world entities, after all, have no Section 230 at all, let alone a procedurally efficient version. That includes retailers, product manufacturers, and mass media distributors and publishers, who, like many online entities, transact business with and distribute content to millions of individuals every day. Even if Section 230 were removed altogether, online entities would still have available all the traditional defenses of physical-world entities. Thus, were an immunity defense more costly to assert, an internet entity might, before raising Section 230, first assert a lack of proximate cause,239 lack of legal duty,240 or a

---

236 *See id.* at 36–37.

237 *See id.* at 39–44.

238 *See* Novak v. Overture Servs., Inc., 309 F. Supp. 2d 446, 452 (E.D.N.Y. 2004) ("Section 230[ ] immunity constitutes an affirmative defense . . . [that] is generally not fodder for a Rule 12(b)(6) motion [to dismiss, but] is generally addressed as [a] Rule 12(c) [post-answer motion for judgment on the pleadings] or Rule 56 motion [for summary judgment]."); Goldman, *supra* note 6, at 39 ("Often, judges can resolve all three elements based solely on the allegations in the plaintiff's complaint. Thus, courts can, and frequently do, grant motions to dismiss based on a Section 230(c)(1) defense.").

239 Even with Section 230 in its current form, courts frequently resolve cases against online entities for lack of proximate cause rather than Section 230. *See, e.g.*, Crosby v. Twitter, Inc., 921 F.3d 617, 623–27, 627 n.7 (6th Cir. 2019) (dismissing claims against platforms alleging liability for Florida nightclub attack for lack of proximate causation without reaching Section 230 issue); Fields v. Twitter, Inc., 881 F.3d 739, 743–50 (9th Cir. 2018) (dismissing suit alleging Twitter knowingly provided material support to terrorist group because harms suffered by plaintiffs lacked direct relationship with Twitter's provision of a messaging platform); *see also* Vesely v. Armslist, LLC, No. 13 CV 00607, 2013 WL 12323443, at *2–3 (N.D. Ill. July 29, 2013) (dismissing wrongful death claim against Armslist website without addressing Section 230 defense because plaintiff failed to allege a sufficiently close nexus between the website and the plaintiff killed with illegally purchased firearm), *aff'd*, 762 F.3d 661 (7th Cir. 2014).

240 *See, e.g.*, Maynard v. Snapchat, Inc., No. A20A1218, 2020 WL 6375424, at *3–4 (Ga. Ct. App. Oct. 30, 2020) (dismissing design defect claim against Snapchat based on lack of duty of conduit liability defense,[241] or move to compel arbitration[242] pursuant to its terms of service. Streamlining the internet immunity defense is an important goal. But given physical-world entities' ability to survive without any Section 230 defense at all, adjustments that decrease the defense's procedural efficiency should not be seen as an insurmountable barrier to reform—at least not without some showing that online entities both merit special treatment and are uniquely unable to bear the costs of asserting the legal defenses available to them.[243]

## B. Designing a Modern Immunity Doctrine

Section 230's protections continue to play a critical role for many online entities, and there are reasons for caution in reforming Section 230. But there is no reason that Section 230 must be preserved in exactly its present form. The internet has changed dramatically over the

---

app developer to protect against third-party misuses of product); *Vesely*, 2013 WL 12323443, at *2.

[241] *Cf.* RESTATEMENT (SECOND) OF TORTS § 581 cmt. f (AM. L. INST. 1965) ("One who . . . transmits a written message for another is not liable for the defamatory character of the message unless he knows or has reason to know that the message is libelous."); Skorup & Huddleston, *supra* note 96, at 644–46, 646 n.65 (discussing conduit liability defense's historical relationship to Section 230).

[242] *See* Jeremy B. Merrill, *One-Third of Top Websites Restrict Customers' Right to Sue*, N.Y. TIMES (Oct. 23, 2014) https://www.nytimes.com/2014/10/23/upshot/one-third-of-top-websites-restrict-customers-right-to-sue.html [https://perma.cc/ZEW4-2K2R].

[243] What limited data are available suggests that litigation costs for technology companies are in line with other industries. *Compare* ENGINE, *supra* note 105, at 1–2 (noting Section 230 litigation costs for startups of millions of dollars), *with* Laws. for Civ. Just., Civ. Just. Reform Grp., U.S. Chamber Inst. for Legal Reform, Appendix 1: Litigation Cost Survey of Major Companies, Presentation to the Committee on Rules of Practice and Procedure Judicial Conference of the United States 14–15 (May 10–11, 2010), https://www.uscourts.gov/sites/default/files/litigation_cost_survey_of_major_companies_0.pdf [https://perma.cc/3VQJ-TESW] (noting similar litigation costs in other industries). The most in-depth analysis to date is CHRISTIAN M. DIPPON, NERA ECON. CONSULTING, ECONOMIC VALUE OF INTERNET INTERMEDIARIES AND THE ROLE OF LIABILITY PROTECTIONS (2017), https://internetassociation.org/wp-content/uploads/2017/06/Economic-Value-of-Internet-Intermediaries-the-Role-of-Liability-Protections.pdf [https://perma.cc/EFM7-KYPU]. Dippon estimates that weakening Section 230 would eliminate 425,000 jobs in the tech sector and reduce annual GDP by $44 billion. *See id.* at 18. Unfortunately, the estimates predate most current reform proposals and rely on several assumptions that limit their usefulness. First, the estimates do not account for the many cases in which entities stripped of Section 230 protection would succeed on other grounds, like proximate cause. Second, they assume that any changes would not be targeted and would raise litigation costs for entities like Facebook and Google, two of the nation's largest companies, to the same degree as more controversial entities like Backpage or Armslist, which are much less important to the economy. Third, they calculate the average litigation costs that internet companies would incur under a weakened Section 230 based on a blockbuster $41.5 million copyright infringement jury award against MP3tunes. *See id.* at 1–2, 13–15; EMI Christian Music Grp., Inc. v. MP3tunes, LLC, 844 F.3d 79, 88 (2d Cir. 2016).

last twenty years,[244] and so too has the scope of the need for content-moderation immunity. If Section 230 is to be updated, however, changes must be carefully structured to account for the concerns raised in the previous section. Specifically, a revised Section 230 must (1) be flexible enough to govern the wide variety of entities and conduct that now occur in the virtual world; (2) continue to protect online entities against the burden to moderate third-party content, which could lead them to proactively censor their users' speech to mitigate the risk of legal liability; and (3) be designed with procedural efficiency in mind to limit litigation costs incurred by online entities in asserting the defense.

Taking these goals as guideposts for reform, this Article proposes that 47 U.S.C. § 230(c)(1) be revised as follows to narrow Section 230 immunity to only those claims that would impose a content-moderation burden on online entities by requiring them to review and moderate third-party created content:

> No provider or user of an interactive computer service shall be treated as the publisher or speaker of any information provided by another information content provider *in any action in which prevention of the alleged harm would require the provider or user of the interactive computer service to review and moderate information provided by another information content provider*.

With this revision, online entities would continue to enjoy robust immunity against those claims that would undercut freedom of expression by imposing impossible burdens on them to review user content. But by tailoring immunity to only those claims in which it is required to avoid the collateral censorship concern,[245] the reform would free

---

[244] *See supra* Section I.C.3.

[245] Rather than focusing on collateral censorship and the difficulty of moderating content to prevent the alleged harm, some courts consider whether the platform offered "neutral tools" to its users. *See, e.g.*, Jones v. Dirty World Ent. Recordings, LLC, 755 F.3d 398, 410–11 (6th Cir. 2014) (applying neutral tools test and reasoning that a "material contribution to the alleged illegality of the content does not mean merely taking action that is necessary to the display of allegedly illegal content[,] [but] [r]ather, it means being responsible for what makes the displayed content allegedly unlawful"). When the Ninth Circuit created the neutral tools test in *Roommates.com*, it may have been reasoning from first principles that entities should not be liable unless they sufficiently "contribute to [the] alleged illegality," 521 F.3d 1157, 1169 (9th Cir. 2008), or it may have been borrowing from the "material contribution" and "staple article of commerce" doctrines, which protect manufacturers of products with substantial noninfringing uses from indirect copyright infringement claims. *See* 3 MELVILLE B. NIMMER & DAVID NIMMER, NIMMER ON COPYRIGHT § 12.04 (1991) (discussing limitations on indirect copyright liability). Either way, the reasoning is inappropriate when evaluating an affirmative defense under Section 230. The statute does not bar claims against insufficient contributors, only claims that

the courts to treat online and offline entities equally in other contexts, including some of the cases discussed above involving product liability, online marketplaces, and volitional wrongdoing.[246] This approach would also account for many of the risks and limitations that have hindered reform efforts.

### 1. Flexibility for a National Problem

One significant obstacle to internet immunity reform is the extreme breadth of the modern internet. The virtual world now encompasses nearly every variety of human conduct, making it difficult to develop a single rule to govern it.[247] Such a law must handle everything from publication-centric claims like defamation to other tort claims sounding in negligence, fraud, or product liability, and civil claims involving antitrust or criminal statutes. Indeed, to be truly comprehensive, internet immunity doctrine must account for every possible claim in the collective statutory and common law of the federal government and all fifty states.

Given the enormity of this task, many reform proposals have focused on narrow carve outs that correct Section 230's treatment of one category of especially problematic claims, for example, intentional facilitation of criminal wrongdoing.[248] Another possibility is to invoke an extremely flexible principle, such as reasonableness, that courts or a regulatory body can update with the times and tailor to suit particular contexts.[249] As discussed above, each of these approaches has advantages and limitations.[250]

A third possibility is that, given the need for flexibility in this area, control over the question of online intermediary liability should be returned to the state common law system. Common law decision making has the advantage of greater flexibility over statutory law.[251] And a common-law system would have the ability to adapt with the internet as it continues to evolve in the coming years. Moreover, as

---

would impose a content-moderation burden on an online entity by treating it as a publisher or speaker. Where those concerns are not implicated, the sufficiency of the defendant's alleged conduct to support a cognizable claim should be assessed under the third-party liability principles of the relevant state or federal law, not Section 230.

246  *See supra* Section II.B.
247  *See supra* Section III.A.2.
248  *See, e.g.*, DOJ Section 230 Recommendations, *supra* note 22, at 14–18.
249  *See, e.g.*, Citron & Wittes, *supra* note 25, at 419.
250  *See supra* Section III.A.
251  *See* Linkins v. Protestant Episcopal Cathedral Found., 187 F.2d 357, 360–61 (D.C. Cir. 1950) ("[T]he very term 'common law' means a system of law not formalized by legislative action, not solidified but capable of growth and development at the hands of judges.").

discussed earlier[252] and detailed by Brent Skorup and Jennifer Huddleston in their recent article,[253] the Section 230 regime enacted by Congress and interpreted by the courts is not a dramatic departure from the common-law rules that govern traditional publishers and distributors.[254]

The common law's organic adaption to the early internet[255] and its continuing able governance of the physical world of newspapers, books, pamphlets, radio, and television suggest that, with time, internet media could safely be added to its domain and that, as with those media, the law would gradually come to account for the specific and evolving nuances of internet technology. Indeed, that may be exactly how Congress envisioned the law would unfold when it enacted Section 230.[256] Despite the flexibility advantage of a common law system, however, a few factors counsel against such a transition. First, there would be the usual upfront costs of switching to a new regime. Internet entities would need to assess the new law's impact on their existing and future operations, and courts would be tasked with creating a new body of decisional law. These latter costs would be unusually high, as courts would need to instantly account for current technology despite a decades-long gap in the case law.

Second, returning internet immunity law to the state courts would forfeit the benefits of a uniform legal framework to govern an important, interstate issue.[257] Even now, of course, jurisdictional splits over the interpretation of Section 230 are possible, but in practice courts across the country have converged on most issues, and there are relatively few differences between jurisdictions. Were the question governed by state common law rather than federal statutory law, the economic costs of multijurisdictional compliance would be significant.

---

[252] *See supra* Section I.C.2.

[253] Skorup & Huddleston, *supra* note 96.

[254] *See id.* at 637–46.

[255] *See* KOSSEFF, *supra* note 12, at 36–56 (discussing courts' capable, pre–Section 230 common-law analysis of online intermediary liability).

[256] Section 230 came into being as part of the CDA and was intended primarily—and, arguably, exclusively—to encourage self-censorship through Section 230(c)(2)'s grant of absolute immunity to entities who engaged in good-faith censorship efforts. *See* 47 U.S.C. § 230(b). If, as some argue was Congress's intent, Section 230(c)(2) had been interpreted as the *only* operative portion of Section 230, all instances of non–self-censorship—that is, the majority of the modern internet—would have continued to be governed by print-media common law. *See* Doe v. GTE Corp., 347 F.3d 655, 660 (7th Cir. 2003) (offering in dicta that perhaps Section 230(c)(1) should be read as a definitional clause and only Section 230(c)(2)'s Good Samaritan provision be read to confer immunity); Citron & Wittes, *supra* note 25, at 403 (interpreting Section 230's legislative history to support only Good Samaritan immunity, not broad intermediary immunity).

[257] The question is similar to the ongoing debate regarding whether privacy law should be governed at the state or federal level. California's adoption of the California Consumer Privacy Act, Cal. Civ. Code §§ 1798.100–.199 (West 2020), which applies even to entities outside of California's borders, has spurred calls for the federal government to enact federal privacy legislation to preempt state law on the subject and create a uniform national standard. *See* Tony Romm, *The Lobbying War Over California's Landmark Privacy Law*, WASH. POST, Feb. 9, 2019, at A15.

Recognizing this problem, this Article's approach would leave in place a federal statutory immunity regime to set a baseline immunity defense for online entities against all state or federal claims that implicate content moderation. As they are now, states would be free to provide online entities *more* protection from claims brought under their laws, but as a preemptive federal defense, Section 230 would continue to provide a baseline level of immunity.

To build flexibility into Section 230, while still providing baseline federal protections, the proposed approach avoids relying on particular mental states or categories of wrongs and instead trims the scope of immunity to its bare essentials—claims requiring content moderation and risking collateral censorship. By allowing all other claims to proceed freely, this approach can rely on the existing body of state and federal law to fill in the requisite mental states and elements required for a plaintiff to make out a claim, whether it be negligence, product liability, civil conspiracy, or something else.

### 2.  *Moderation and Collateral Censorship*

This proposal also avoids interfering with the content-moderation and collateral-censorship protections that Section 230 was designed to create. To capture the most egregious wrongdoing by online entities, such as Backpage's allegedly intentional facilitation of child sex trafficking, some reform proposals have suggested limiting immunity in cases that involve intentional or knowing wrongdoing.[258] Although these proposals might achieve their goal, there is a risk that adding a scienter component to the Section 230 analysis would recreate the collateral censorship concern. Plaintiffs could plead intent or knowledge without factual support, and online entities might preemptively censor user speech or immediately remove all flagged content to avoid that risk.[259]

Rather than inquire into a defendant's mental state, however, this proposal would instead tie immunity to an objective rule. Online entities would be immune from any claim that would require them to review and moderate third-party-created content. Far from exacerbating the collateral censorship concern, this proposal would reduce it, by introducing express language into the statute to guard against any judicial interpretation that might read Section 230 to permit a claim predicated[260] on a failure to censor third-party speech.

---

258  *See supra* Section III.A.1.
259  *See* Goldman, *supra* note 6, at 38, 45.

### 3. Limiting Litigation Costs

Finally, this proposal would update Section 230 without dramatically increasing the cost of asserting the defense in litigation. Under the current regime, when a defendant asserts a Section 230 defense, courts are typically able to resolve that defense by pretrial motion where it is clear from the complaint that there is no dispute that the plaintiff's claim would make an online entity responsible for content authored by a third party.[261] Similarly, even where there is some allegation that the defendant was a cocreator of the content at issue, courts are often able to decide Section 230 issues pretrial on a motion for summary judgment following limited discovery.[262] Were Section 230 modified to incorporate a mental state analysis or compliance with a specified standard of care, this procedural benefit would be lost.

To avoid unnecessarily diminishing Section 230's procedural benefits, the proposed approach relies on an objective test that will often be resolvable on the pleadings alone or following limited discovery. Just as, under the current rule, it is often clear on the face of a complaint whether the defendant online entity, rather than some third party, authored the content in question, under the proposed rule it would often be clear whether preventing the complained of harm would have required the entity to review and moderate content created by another party.

---

[260] Sometimes multiple approaches will be available to prevent complained of harms, some of which impose a content-moderation burden and others of which do not. In such cases, Section 230 would bar only those theories that would impose a content-moderation burden. For example, were an entity to develop an app to connect medical patients with third-party physicians and host virtual appointments, a patient harmed by errant medical guidance could assert a negligence claim against the entity for failing to verify her physician's license to practice medicine, but not for failing to review the quality of medical guidance provided. The reasonableness of the entity's practices regarding validation of participant physicians' credentials would then be assessed under state tort law.

[261] *See, e.g.*, E. Coast Test Prep, LLC v. Allnurses.com, Inc., 971 F.3d 747, 752 (8th Cir. 2020) (affirming grant of motion for judgment on the pleadings under Section 230 in action brought by test prep company against platform that hosted negative user reviews despite plaintiff's implausible allegation that the platform had intentionally solicited the negative reviews).

[262] *See, e.g.*, Frontier Van Lines Moving & Storage, Inc. v. Valley Sols., Inc., No. 11CV0526, 2011 WL 2110825, at *5 (W.D. Pa. May 24, 2011) (ordering limited discovery on Section 230 immunity issue "out of an abundance of caution"); *see also* Novak v. Overture Servs., Inc., 309 F. Supp. 2d 446, 452 (E.D.N.Y. 2004) (explaining that Section 230 immunity is "generally addressed as a Rule 12(c) [post-answer motion for judgment on the pleadings] or Rule 56 motion [for summary judgment].").

A leading case brought against the online review aggregator Yelp is a good example. In *Kimzey v. Yelp!, Inc.*,[263] the owner of a locksmith business asserted libel, unfair competition, and RICO[264] claims against Yelp based on two negative reviews that were posted about his business on the platform.[265] The plaintiff argued that their business had been harmed by the posts and that Yelp was responsible for those harms because it created the star-based rating metric used by its reviewers and caused the statements to appear not only on its own site, but also as part of a promotion on Google's search engine.[266] Because the claim predicated liability on content authored by a third-party, the court affirmed the dismissal of the action under Section 230 for failure to state a cognizable claim.[267]

The result would be the same under the proposed reform. Yelp acts as an aggregator and passthrough for customers to write and post reviews about their experiences with businesses.[268] To prevent the harm of which the *Kimzey* plaintiff complained, Yelp would need to review and determine the truth or falsity of the third-party reviews hosted on its platform. A court could handily resolve such a claim on a motion to dismiss under either standard.

### C.  *Uniting the Physical and Virtual Worlds*

Just as important as doing no harm to the current internet immunity regime are the beneficial changes that this proposal would contribute. As discussed above, Section 230 currently privileges online entities over offline ones by providing them with a special defense unavailable to their physical-world counterparts.[269] By narrowly tailoring internet immunity to only those contexts in which it is necessary to avoid the collateral censorship concern, the proposed revision would reduce the disparity in the law's treatment of online and offline entities by allowing courts to treat the two alike in other contexts.

---

[263] 836 F.3d 1263 (9th Cir. 2016).
[264] Racketeer Influenced and Corrupt Organizations ("RICO") Act, 18 U.S.C. §§ 1961–1968.
[265] *See Kimzey*, 836 F.3d at 1265–67.
[266] *Id.* at 1266.
[267] *Id.* at 1266, 1271.
[268] *About Yelp*, YELP, https://www.yelp.com/about [https://perma.cc/38LJ-6V36].
[269] *See supra* Sections I.B–.C.

For example, recall the earlier discussion of the Snapchat case and product liability claims involving websites or smartphone apps.[270] Whereas in the physical world, a party injured by a defective product can recover via a negligence or product liability action, that option may be unavailable to a plaintiff harmed by a virtual world product, whose creator can assert immunity under Section 230.[271] That is exactly what happened in *Lemmon*, where the court dismissed the plaintiff's claim against Snapchat for negligent design of the speed filter component of its picture-taking app—which contributed to numerous car crashes—on the ground that Snapchat was not the author of any content.[272]

Contrast that result with how a court would approach the Snapchat immunity question under the proposed reform. Rather than ask by whom content is authored, a court would ask whether preventing the harms suffered by the plaintiff would have required Snapchat to review and moderate information provided by a third party. In *Lemmon*, the answer was no. The plaintiff alleged not that Snapchat was negligent in failing to identify and censor damaging communications, but that the speed filter component of its app was defectively designed.[273] Applying the suggested reform, a court would deny Section 230 immunity and consider the cognizability of the claim under state negligence or product liability law, just as it would with a physical-world product.

The same would be true for many claims against online marketplaces, another context where Section 230 currently creates a disparity between online and offline entities.[274] Consider the *Backpage.com* decision.[275] In that case, a group of three underage girls who had been advertised for sale on the Backpage website brought a statutory sex-trafficking conspiracy claim[276] under the federal TVPRA.[277] Even though the plaintiffs alleged that Backpage had designed its website to facilitate and profit from sex trafficking and that it had taken specific, intentional actions to thwart law-enforcement efforts (by removing posts connected with police sting operations and wiping metadata from escort photos), Section 230 immunized Backpage from liability

---

270 *See supra* Section II.A.
271 *See* Lemmon v. Snap, Inc., 440 F. Supp. 3d 1103, 1105 (C.D. Cal. 2020).
272 *See id.* at 1105–07.
273 *See Lemmon*, 440 F. Supp. 3d at 1106.
274 *See supra* Section II.B.
275 Doe v. Backpage.com, LLC, 817 F.3d 12 (1st Cir. 2016).
276 *See* Backpage Complaint, *supra* note 163, at ¶¶ 108–14.
277 22 U.S.C. §§ 7101–7114.

because the sex traffickers, not Backpage, authored the online listings.[278]

Under the proposed reform, however, the plaintiffs would have been free to pursue their claim against Backpage because they alleged that Backpage participated in the sex-trafficking conspiracy not by failing to review and remove offending posts, but by implementing platform features designed to profit from sex trafficking and avoid detection by law enforcement. Applying the suggested reform, a court would have denied Section 230 immunity and proceeded to consider the claim under the TVPRA.

Finally, it is important to discuss a few key scenarios that would, and should, remain unaffected by the proposed reform. Recall the earlier discussion of cases involving volitional wrongs, where an online entity's conduct is wrongful because of some knowledge of or even intent to cause harms.[279] Because Section 230 looks only to content authorship, an online entity can be immune from liability even in cases of volitional wrongdoing.

Although it may be tempting to withhold immunity in such contexts, doing so could undermine Section 230's key objectives. Consider the content-flagging mechanisms available on many apps and online platforms, which give users an easy way to notify the online entity of problematic content. Content-flagging mechanisms are a good thing. They make it easier for users to report and easier for entities to find and remove unlawful content. But they do not make it costless. Indeed, online platforms receive so many alerts that reviewing the flagged content itself poses a significant moderation burden, despite the knowledge of the content and its potentially problematic character.[280] Knowledge-based liability could lead to collateral censorship, as platforms might take down any content reported as objectionable, without investigation, to avoid the risk of liability or the burden of moderating content.[281]

Claims of volitional wrongdoing would be allowed to proceed, however, where they would not require an online entity to review and moderate content. The *Barnes* case is a good example. In that case, the plaintiff's former boyfriend created fake online profiles under her name in order to induce strangers to proposition her for sex.[282] Under the proposed reform, Yahoo would be immune from claims related to its failure to detect and remove the fake profiles, even once those profiles were flagged. But after Yahoo contacted the plaintiff to tell her it would remove the profiles,[283] it could be liable for its continuing failure to do so. Following through on its promise to remove a piece of content would not require additional content moderation.

---

278  *See Backpage.com*, 817 F.3d at 16–17, 22.

279  *See supra* Section II.B.

280  *See* Naughton, *supra* note 202; *see also* Goldman, *supra* note 6, at 38–41 (explaining that a knowledge-based liability rule is equivalent to a notice-and-takedown rule because the economically rational response to complaints would be to remove the challenged content without investigation).

281  *See* Wu, *supra* note 6, at 295–96.

The proposed reform would thus bring the online and offline world into closer, if not complete, alignment. Tailoring internet immunity doctrine narrowly to address its core content moderation and collateral censorship concerns would update Section 230 to account for the internet's evolution over the last twenty years, while still protecting against the very real concerns that motivated the statute's enactment.

## Conclusion

Current immunity doctrine is based on an outdated, mass-media-inspired understanding of the internet that is ill-suited to govern the diverse collection of entities that populates the online world today. Rather than merely relay information and communications, modern websites allow people to do everything from romantic matchmaking to coordinating transportation networks, booking vacation packages, and selling used kitchen appliances. Section 230's outdated conception of the internet has created a disparity between the law's treatment of online versus offline entities and produced calls for reform across the political spectrum. This Article lays the groundwork for Congress and the courts to reform internet immunity doctrine, bring the law's treatment of the virtual and physical worlds into alignment, and treat like cases alike wherever they occur.

---

282 *See* Barnes v. Yahoo!, Inc., 570 F.3d 1096, 1098–99 (9th Cir. 2009).
283 *See id.*